\input harvmac
\baselineskip=.55truecm
\Title{\vbox{\hbox{HUTP--95/A037}\hbox{hep-ph/9603444}}}
{\vbox{\centerline{Form factor $\pi^0\to \gamma^* +\gamma^*$ at different photon virtualities}}}
\vskip .1in
\centerline{\sl A. Anselm$^{a,}$\foot{E-mail: anselm@thd.pnpi.spb.ru}, 
A. Johansen$^{a,b,}$\foot{E-mail: johansen@string.harvard.edu}, 
E. Leader$^{c,}$\foot{E-mail: e.leader@physics.bbk.ac.uk} and \L.
\L ukaszuk$^{d,}$\foot{E-mail: lukaszuk@fuw.edu.pl}}
\vskip .2in
\centerline{$^a$\it The St.Petersburg Nuclear Physics Institute, Gatchina, 188350, Russia}
%\vskip .1in
\centerline{$^b$ \it 
Lyman Laboratory of Physics, Harvard University, Cambridge, MA 02138, USA}
%\vskip .1in
\centerline{$^c$\it Birkbeck Colledge, University of London, 
Malet Street, London WC1E 7HX, England}
%\vskip .1in
\centerline{$^d$\it Soltan Institute for Nuclear Studies, Hoza 69, 
00-681 Warsaw, Poland}
\vskip .2in
%\draft
\centerline{ABSTRACT}
\vskip .2in
The $\pi^0 \gamma\gamma$ vertex for virtual photons of squared masses
$q_1^2$ and $q_2^2$ plays a vital r\^ole in several physical processes;
for example for $q_1^2<0$, $q_2^2<0$, in the two-photon physics reaction
$e^+ e^-\to e^+ e^- \pi^0$, and for $q_1^2>0$, $q_2^2>0$,
in the annihilation process  $e^+ e^-\to  \pi^0 l^+ l^-$.
It is also of interest because of its link to the axial anomaly.
We suggest a new approach to this problem.
We have obtained a closed analytic expression for the vertex in the 
limit in which at least one of $|q_1^2|$ and $|q_2^2|$ is large
for arbitrary fixed values of the ratio $q_1^2/q_2^2$.
We compare our results with those obtained previously by Brodsky and Lepage.
It should be straightforward to test our predictions experimentally.

\Date{November 1995}

\newsec{Introduction}

As is well known the famous Adler-Bell-Jackiw anomaly leads to a non-vanishing amplitude
 for neutral pion decay, $\pi^0\to 2\gamma$, in the chiral limit when the pion mass and 
the current masses of light quarks go to zero: $m_{\pi}\to 0$, $m_u ,m_d\to 0$.
The result is a consequence of the linear divergence of the triangle diagram and the
 whole contribution actually comes from short distances when a gauge invariant 
regularization is utilized.
A natural question arises: does anything change when the photons are off-shell?
At first glance, since the anomalous contribution has a point-like 
character, nothing 
changes seriously provided the virtualities (masses) of photons 
remain smaller than the 
scale of regularization, which, however, can be chosen arbitrarily large. 
This simple observation 
led Jacob and Wu some years ago \ref\wu{M. Jacob and T.T. Wu, Phys. Lett. {\bf B 232} 
(1989) 529.} to very strong statements about certain physical processes. 
For instance, if 
the decay amplitude $Z\to \pi^0 +\gamma$, which is intimately related to $\pi^0\to \gamma^*+
\gamma$, is controlled by a point-like amplitude and does not decrease as $m(\gamma^*)
\to m_Z$, one has a $\pi^0 \gamma$ production rate wildly incompatible with LEP data.
The above expectation seems, of course, quite unintuitive {\it per se}; 
still stranger that  at the time of publication of ref. \wu \
the correct answer had been known \ref\marc{L. Arnellos, W.J. Marciano and Z. Parsa,
Nucl. Phys. {\bf B 196} (1982) 371.} for some years. 
In fact an extremely small value of the amplitude of 
$Z^0\to\pi^0 +\gamma$ decay obtains because of the decrease of the amplitude 
$\pi^0\to \gamma+\gamma^*$ as the virtuality of the photon increases
\ref\brod{S.J. Brodsky and G.P. Lepage, Preprint SLAC-PUB-2294 (March 1979);
Phys. Lett. {\bf B 87} (1979) 359; Phys. Rev. {\bf D 22} (1980) 2157; 
{\it ibid.} {\bf D 24} (1980) 1808.}
\ref\rad{A.V. Radyushkin, Preprint P2-10717 (June 1977)\semi
A.V. Efremov and A.V. Radyushkin, Preprint E2-11535 (April 1978)\semi
A.V. Efremov and A.V. Radyushkin, Preprint E2-11983 (October 1978), 
Theor. Math. Phys. {\bf 42} (1980) 97.}
\ref\zh{V.L. Chernyak, A.R. Zhitnitsky and V.G. Serbo,
JETP Lett. {\bf 26} (1977) 594.}.
Another similar amplitude, which had also been considered earlier, is $\pi^0\to \gamma^*+
\gamma^*$ with equal, large negative virtualities for the photons, $q_1^2=q^2_2=-Q^2$ 
\ref\volo{M.B. Voloshin, preprint ITEP-8, 1982 (unpublished)\semi
V.A. Novikov, M.A. Shifman, A.I. Vainshtein, M.B. Voloshin and V.I. Zakharov,
~~~~Nucl. Phys. {\bf B 237} (1984) 525\semi
V.A. Nesterenko and A.V. Radyushkin, Yad.  Fiz.  {\bf 38} (1983) 476
(Soviet J. of Nucl. Phys. {\bf 38} (1983) 284).}.
For this case a simple operator product expansion (OPE) leads to the natural conclusion 
that the amplitude falls as $1/Q^2$ in the asymptotic region $Q^2\to \infty$.

The paper \wu \ has already been criticized \ref\phua{T.N. Pham and X.Y. Pham,
Phys. Lett. {\bf 247 B} (1990) 438.}
but since to some extent it triggered this investigation we shall give in the next section 
a simple explanation of the ``paradox'' described above.

There are interesting experimental motivations to study the amplitude 
$\pi^0\to \gamma^*+\gamma^*$ for non-vanishing masses 
$q_1^2\neq 0,\;\;\; q_2^2\neq 0$ of  the virtual photons.
This amplitude can be measured in principle as a function of negative 
$q_1^2< 0$ and $q_2^2< 0$ in ``two-photon physics'' (see Fig.1).
For positive values $q_1^2,\; q_2^2> 0$ it enters the amplitude of the process $e^+ +
e^-\to \pi^0 +l^+ +l^-$ (Fig.2).
Though this process is 
too small to be measured when it goes through the $Z$ boson it 
can be observed in $e^+ e^-$ collisions at lower energies.

Consequently the interest in the amplitude $\pi^0\to \gamma^*+\gamma^*$ 
is far from exhausted 
and there is a growing literature on the subject \ref\bulk{F. Del Auila and 
M.K. Chase, Nucl. Phys. {\bf B 193} (1981) 517\semi
E.P. Kadantseva, S.V. Mikhailov and A.V. Radyushkin, 
Yad. Fiz. {\bf 44} (1986) 507 (Soviet J. of Nucl. Phys. {\bf 44} (1986) 326)\semi
J. West, Mod. Phys. Lett. {\bf A5} (1990) 2281\semi
A.V. Radyushkin and R. Ruskov, Phys. Atom. Nucl.
{\bf 56} (1993) 603\semi
A.V. Radyushkin and R. Ruskov, Phys. Atom. Nucl. {\bf 58} (1995) 1356\semi
R. Jakob, P. Kroll and M. Raulfs, preprint WU B 94-28\semi
A.V. Radyushkin and R. Ruskov, `QCD Sum Rule Calculation of
$\gamma\gamma^*\to \pi^0$ Transition Form Factor', CEBAF-TH-95-17, 
hep-ph/9511270; `Transition Form Factor $\gamma \gamma^*\to \pi_0$ and QCD 
Sum Rules', CEBAF-TH-95-18, hep-ph/9603408;
`Form Factor of the Process $\gamma^*\gamma^*\to \pi^0$ for Small
Virtuality of One of the Photons and QCD Sum Rules,'
Contribution to the PHOTON95 Conference, Sheffield (1995),
hep-ph/9507447, and references therein\semi
V.V. Anisovich, D.I. Melikhov and V.A. Nikonov,
`Photon-meson transition form factors $\gamma\pi^0$, $\gamma\eta$ and
$\gamma\eta'$ at low and moderately high $Q^2$', hep-ph/9607215.}, 
\ref\gorsky{A.S. Gorsky, Yad. Fiz. {\bf 46} (1987) 938
(Soviet J. of Nucl. Phys. {\bf 46} (1987) 537).},
\ref\mikrad{S.V. Mikhailov and A.V. Radyushkin, Yad. Fiz.  
{\bf 52} (1990) 1095 (Soviet J. of Nucl. Phys. {\bf 52} (1990) 697).}.
Indeed experiments in ``two-photon physics'' have now been started by CLEO
at Cornell.
For a summary of data up to the end of 1994 see ref. \ref\morgan{D. Morgan, M.R. 
Pennington and M.R. Whalley, J. Phys. G: Nucl. Part. Phys. {\bf 20} (1994) A1.}.

The Feynman amplitude for $\pi^0 (k)\to \gamma^* (q_1)+\gamma^* (-q_2)$
can be written in the form
\eqn\am{{\cal M}=M(q_1^2 ,q_2^2) \cdot 
\epsilon_{\mu\nu\lambda\sigma}\epsilon^{\mu}
(q_1)\epsilon^{\nu}(-q_2)q^{\lambda} k^{\sigma}}
where
\eqn\dfn{q=(q_1+q_2)/2,\;\;\;k=q_1-q_2,\;\;\;{\rm and}\;\;\; q^2=(q_1^2 +q_2^2)/2 -
m_{\pi}^2/4 .}
Two results, of great importance, have been known for many years.
In the case of equal space-like momenta, $q_1^2 =q_2^2<0$ the operator product expansion 
(OPE) yields the following behaviour as $q_1^2 =q_2^2\to -\infty$
\eqn\ope{M(q^2 ,q^2)=
M(0,0)\left(-{8\pi^2 f_{\pi}^2\over 3q^2}\right)\;\;\; {\rm for} 
\;\;\;q^2\to -\infty}
where we have used $q^2\approx q^2_1=q_2^2$ from eqs.\dfn \ and where 
$f_{\pi}\approx 93 MeV$. The amplitude $M(0,0)$ is the usual amplitude 
for $\pi^0\to \gamma+\gamma$ into real photons:
\eqn\real{M(0,0)={e^2\over4\pi^2}\cdot {1\over f_{\pi}}.}
For the case of one photon on mass-shell ($q^2_2=0$), and the other with 
large positive virtuality ($q_1^2>0$), Brodsky and Lepage have used the 
QCD infinite momentum frame approach for the description of the pion 
wave function and have obtained the asymptotic behaviour
\eqn\asy{M(2q^2,0)=M(0,0) \left({4\pi^2f_{\pi}^2\over -q^2}\right) \;\;\; {\rm 
for}\;\;\; q^2\to \infty}
where we have used $q^2\approx q_1^2/2$ from eqs.\dfn .

Generally the amplitude $M$ is a function of the two variables $q_1^2$ 
and $q_2^2$.
It is interesting to consider the case when $q_1^2$
and $q_2^2$ both go to infinity, but their ratio remains finite.
As far as we know for this case no reliable results exist.

In this paper we try to derive some conclusions about the asymptotics in 
the latter region, attacking the problem by means of a triangle diagram 
in which the pion-quark-antiquark vertex is given in its most general 
(non-pointlike) form.
In the next section we shall argue that this diagram, though 
mixing hadron and quark degrees of freedom, gives reliable results for 
our purposes.
This is not a trivial point and the straightforward analogy with the 
triangle diagram always used for $\pi^0$ decay into real photons is 
somewhat misleading.
In the latter case one actually {\it does not} use a diagram containing 
the pion and the quarks, but rather a diagram containing the axial 
current.
It is only through PCAC that we can connect it to $\pi^0\to 2\gamma$ 
decay.
The axial current vertex is pointlike {\it by definition}, quite different 
from the case of the pion vertex which is considered below.

Our results are the following.
First we have tested our approach by reproducing 
the OPE asymptotics for the case of 
large equal and negative virtualities, $q_1^2=q^2_2<0$, eqs. \ope , ~\real .

Second, for the more general case of
a fixed ratio $q_1^2/q_2^2$, but $q_1^2,q_2^2$ going to infinity
we find a ``scaling law'':
\eqn\scaling{M(q_1^2,q_2^2)=M(0,0) \left(-{8\pi^2f_{\pi}^2\over 3q^2}\right)
\phi(\omega),}
$$\omega={q_1^2-q_2^2\over q_1^2+q_2^2}$$
where $\phi(\omega)$ is a certain dimensionless function.

Note that the factor $f_{\pi}^2/q^2$ appears naturally in the 
particular cases described by eqs. \ope \  and \asy .
The variable $\omega$ lies in the interval $-1\leq \omega \leq 1$ 
since we consider the physically interesting cases of either both 
$q^2$'s negative, $q_1^2,q_2^2<0$, (Fig.1) or both positive, 
$q_1^2,q_2^2>0$ (Fig.2).
Eqs. \ope \ and \asy \ imply that 
\eqn\bound{\phi (0)=1,\;\;\;\phi (1)=3/2.}
The principle aim of this paper will be to calculate $\phi (\omega)$ 
for all $-1\leq \omega \leq 1$.
We shall be able to find $\phi (\omega)$ in the chiral limit, 
$m_{\pi}\to 0$, and under some natural physical assumptions.
Our result for $\phi (\omega)$ reads
\eqn\fsc{\phi (\omega)={1\over \omega^2 }[(1+\omega){\rm ln}(1+\omega)+
(1-\omega){\rm ln}(1-\omega)].}
At $\omega=0$ $\phi (0)=1$, as required by eq.\bound ; at $\omega=1$
$\phi (1)=2\; {\rm ln} \; 2\approx 1.4.$
The latter value is slightly different from that given by eq. \bound \ 
which is not surprising since our approach differs from that of ref. \brod .

In fact the form \scaling \ is not new (see, for example, \gorsky).
Generally speaking the exact form of the function $\phi (\omega)$
is determined by the detailed form of the pion wave function.
Our calculation shows,
however, that one can obtain an approximate expression for
$\phi (\omega)$ without an explicit knowledge of the pion wave function,
provided one assumes reasonable asymptotic properties of the wave
function or the related form factors.
Actually in our computation we use an expansion with respect to a formal parameter
related to the asymptotic behaviour of pion wave function. 
The leading behaviour of the form factor with respect to this formal parameter turns out to 
be independent on details of the pion wave function.
In ref. \gorsky \ the validity of the treatment is restricted to the region of
small $|\omega|\leq 0.5$ (for a further analysis in the sum rule approach see
\mikrad).
In ref. \ref\manohar{A. Manohar, Phys. Lett. {\bf B 244} (1990) 101.}
a series for $\phi (\omega)$
is obtained in which succeeding terms are suppressed by powers of
$\log Q^2$ but it is claimed that the result is only reliable for small omega.
Contrary to the approaches of refs. \gorsky \ and
\manohar \ we use a numerical approximation, which,
we hope, may be valid for $|\omega|\leq 1$.
Note also that our treatment ignores corrections of order
$1/\log Q^2$.
Therefore our result may be formally interpreted as an expression
for the amplitude in the limit of very large $Q$ when the corrections
$\sim 1/\log Q^2$ can be neglected.
However since we exploit a quite different physics as compared to
ref. \gorsky \ we think that our result is a good approximation
also for `intermediate' $Q$ which are larger than typical
hadronic scale but $\log Q^2$ is not negligible.

In the last part of the paper we apply our formalism to the $\pi^0$ decay
into real photons.
We calculate the decay amplitude and compare it with the PCAC result.
This leads to a relation between one of the form factors describing 
the pion-quark-antiquark vertex at small momenta and $f_{\pi}$.
The implication of this relation is that at small momenta the effective 
pion-quark-antiquark Lagrangian has the form
\eqn\effL{L_{eff}= 
{1\over f_{\pi}} (\partial_{\mu}\Phi_{\pi}) (\bar{q}
\tau_3\gamma^{\mu}\gamma_5 q).}
The paper is organized as follows. In section 2 we discuss the 
Jacob-Wu ``paradox'' and introduce the main formalism which we use.
In section 3 we calculate the constant $f_{\pi}$ in terms of integrals 
over the introduced form factors and study the asymptotics of $\pi^0 
\to \gamma^* +\gamma^*$ at equal large photon masses $q_1^2=q_2^2=q^2\to \pm \infty .$
The OPE result for this asymptotics is reproduced.
In section 4 we calculate the function $\phi (\omega)$, eq. \fsc .
In section 5 we deal with $\pi^0$ decay into real photons and derive 
the relation between the zero momentum value of one of our form factors and
$f_{\pi}$, corresponding to eq. \effL .
In the conclusion we summarize and  discuss briefly our results.

\newsec{General formalism}

To set the stage we begin by discussing the Jacob-Wu ``paradox'' in which 
the $\pi^0\to \gamma^*+\gamma^*$ amplitude does not decrease with the large photon 
virtualities.

In considering $\pi^0$ decay into two real photons one uses 
the PCAC relation and the divergence of the axial-vector current $A^{\mu}$,
corrected by the Adler-Bell-Jackiw anomalous term
\eqn\anom{\partial_{\mu} A^{\mu}=f_{\pi} m_{\pi}^2 \Phi_{\pi}
+{e^2\over 16\pi^2} F_{\mu\nu}\tilde{F}^{\mu\nu},\;\;
A_{\mu}=\bar{q}{\tau_3\over 2}\gamma_{\mu}\gamma_5 q}
where the notation is standard.

From eq.\anom \ one gets, by well-known arguments, for the $\pi\to
2\gamma$ amplitude
\eqn\pc{{\cal M}={1\over f_{\pi}} <2\gamma|\partial_{\mu}A^{\mu}|
0>+{1\over f_{\pi}}\cdot {e^2\over 4\pi^2} \cdot 
\epsilon_{\mu\nu\lambda\sigma} \epsilon^{\mu}(q_1)
\epsilon^{\nu} (-q_2)
q^{\lambda}k^{\sigma}.}
The matrix element of $\partial_{\mu}A^{\mu}$ between the vacuum
 and $2\gamma$ state is proportional to $k^2$ where $k$ is the
 four momentum of the pion and of the $2\gamma$ state. 
Since there are no relevant physical massless particles 
in the limit $k_{\mu}\to 0$ the first term in eq.\pc \ 
vanishes and we remain with the second term which is the famous 
result for the $\pi\to
2\gamma$ amplitude.
Why is this not correct when the $\gamma$s are off shell?
After all, as has already been mentioned, the contribution 
proportional to $F_{\mu\nu}\tilde{F}^{\mu\nu}$ comes 
from very short distances and therefore can not change unless the masses of the 
photons reach the regularization scale, which is arbitrarily large.

To understand the changes appearing when the virtualities 
of the photons are not zero, $q_1^2\neq 0$, $q_2^2\neq 0$, 
let us go back one step and write down the full expression
for $<2\gamma|\partial_{\mu}A^{\mu}|
0>$ in eq. \pc , for light but not completely massless quarks,
$m_q=m_u=m_d\neq 0$.
One has~\ref\adler{S. Adler, Phys. Rev. {\bf 177} (1969) 2426.} 
$$\;$$
\eqn\adler{
<2\gamma|\partial_{\mu}A^{\mu}|0>=-{e^2\over 
4\pi^2} \cdot \epsilon_{\mu\nu\lambda\sigma} \epsilon^{\mu}(q_1)
\epsilon^{\nu} (-q_2)
q^{\lambda}k^{\sigma}\times }
$$\times\left[1+2m_q^2 \int^1_0 dx\int_0^{1-x} dy \;\;
{1\over (q_1^2 x+q_2^2 y)(1-x-y) +xy m_{\pi}^2 -m_q^2}\right].$$
The first term in this equation (unity in the parenthesis) 
corresponds to the short distance contribution.
It can be derived, say, by introduction of a Pauli-Villars regulator fermion.
After regularization it is permissible to use the equations 
of motion for the operators in calculating the divergence of $A_{\mu}$.
One then obtains the second term in eq.\adler \ from the convergent triangle diagram 
with light quarks and with a 
pseudoscalar vertex $2im_q \gamma_5$.

Of course, one can not take expression \adler \ too seriously 
since the main contribution to the second term is determined 
by the small momentum domain (of order of $m_q$) of integration.
This corresponds to long distances where the perturbative 
approach is not valid.
However one can use eq.\adler \ to resolve the paradox at least 
at the qualitative level.

Suppose that due to nonperturbative long distance effects 
$m_q$ in eq.\adler \ is a quantity which is closer to a constituent quark mass value rather than to the current quark mass.
Then, in the chiral limit, we can neglect $m_{\pi}^2$ 
in the denominator in eq.\adler .
For the case $q_1^2=q_2^2=0$ the quark mass
$m_q$ cancels out, the integral equals $-1$ and $<2\gamma|\partial_{\mu}A^{\mu}|0>=0.$
According to eq.\pc \ one immediately arrives at the usual result for the $\pi^0\to 2\gamma$ amplitude.
(Note that before we simply argued that matrix element 
$<2\gamma|\partial_{\mu}A^{\mu}|0>$ vanishes due to the absence
of massless physical states.)

For $q_1^2, \; q_2^2\geq m_q^2$ the situation changes drastically.
The second term no longer cancels the unit term in the brackets in
eq.\adler \ and therefore the matrix element of $\partial_{\mu}A^{\mu}$ in eq.\pc \ cannot be neglected.
Moreover, for $q_1^2, \; q_2^2>> m_q^2$ the integral in 
eq.\adler \ is small compared to 1 and 
$<2\gamma^*|\partial_{\mu}A^{\mu}|0>\to -e^2/4\pi^2 \cdot 
\epsilon_{\mu\nu\lambda\sigma} \epsilon^{\mu}\epsilon^{\nu}
q^{\lambda}k^{\sigma} .$
The amplitude ${\cal M} (\pi^0\to 2\gamma^*)$ goes to zero as 
$q_1^2, \; q_2^2\to \infty$ as $\sim 1/q_1^2,$ $\sim 1/q_2^2$,
roughly speaking.

We shall now attempt to attack the problem in a more general
and rigorous fashion.
A key object which we shall use throughout the paper is the
 momentum space proper vertex $\Gamma_{\alpha \beta} (k,l)$ 
describing the transition of $\pi^0$ into a light 
(up- or down-) quark-antiquark pair (Fig.3).
The formal definition of $\Gamma(k,l)$ is as the Fourier
transform of the co-ordinate space vertex 
$\hat{\Gamma}(x_1,x_2)$
\eqn\vertex{i(2\pi)^4 \delta^4 (K-k)\cdot \Gamma(k,l)=\int d^4x_1 \; d^4x_2 \; 
e^{i(l+k/2)x_1 -i(l-k/2)x_2} \hat{\Gamma}(x_1,x_2)=}
$$\int d^4x \; d^4X e^{ikX+ilx} \hat{\Gamma}(x,X),$$
where $X=(x_1+x_2)/2,$ $x=x_1-x_2$, and where the function 
$\hat{\Gamma}(x_1,x_2)$ is defined as the matrix element 
of a product of Heisenberg operators of the quark fields 
between vacuum and pion states. 
For definiteness we use the $u$-quark
\eqn\Gd{\hat{\Gamma}_{\alpha,\beta}(x_1,x_2)
= i \partial\hskip -.085in /  _{\alpha\alpha'}^{(1)} i\partial\hskip -.085in /  _{\beta'\beta}^{(2)} 
<\pi^0|T u_{\alpha'} (x_1) \bar{u}_{\beta'} (x_2)|0>}
where $\alpha$ and $\beta$ are spinor labels; the pion state $<\pi^0|$ is taken with a definite
momentum $K$.
For the $d$-quark the amplitude $\hat{\Gamma}(x_1,x_2)$
is the same but of opposite sign.
The equation \Gd \ is written for a given colour, i.e. a 
summation over colour is not implied.
The vertex $\hat{\Gamma}(x_1,x_2)$ is related to the quark-antiquark component of the wave function of the pion, but 
since we require a proper vertex part we have amputated the
fermion legs in the corresponding Feynman amplitude by applying 
the operators $\partial\hskip -.08in /  ^{(1)}$ and 
$\partial\hskip -.08in /  ^{(2)}$ in eq.\Gd .

Using the definition \Gd \ and invariance under charge 
conjugation one can easily show that for the matrix in spin space
\eqn\charge{\hat{\Gamma}(x_2,x_1)=C^{-1}
\hat{\Gamma}^T(x_1,x_2)C}
where $C=i\gamma^0\gamma^2$ is the charge conjugation matrix.

The most general Lorentz structure of $\Gamma(k,l)$ is
\eqn\gener{\Gamma(k,l)=A\gamma^5+iB\gamma^{\lambda}\gamma^5 
k_{\lambda} +iC (k\cdot l)\gamma^{\lambda}\gamma^5 l_{\lambda}
+D\sigma^{\lambda\mu}\gamma^5 k_{\lambda} l_{\mu} .}
Here the scalar form factors $A,$ $B,$ $C,$ and $D$ depend on two 
scalars, $l^2$ and $k\cdot l$ ($k^2=m_{\pi}^2$).

The relation \charge \ implies that all form factors  $A,$ $B,$ $C$ 
and $D$ are even functions of $l$.
Sometimes it is more convenient to use instead of the scalars 
$l^2$ and $k\cdot l$ the masses of the quark lines
\eqn\mass{m_1^2=(l+k/2)^2,\;\;\; m_2^2=(l-k/2)^2 .}
The eveness under $l\to -l$ implies that $\Gamma (k,l)$ is
invariant under $m_1^2\leftrightarrow m_2^2$.

Before going into the detailed calculations of the asymptotic behaviour of
$\pi^0\to \gamma^*+\gamma^*$ using $\Gamma (k,l)$ we comment on its
relationship to $f_{\pi}$ and on the question of validity of our approach.

Using $q$ for the isospin doublet $(u,d)$, $f_{\pi}$ is defined 
conventionally, by
\eqn\dfnfpi{-if_{\pi} k^{\mu}=<\pi^0 (k)|\bar{q}\gamma^{\mu}
\gamma_5{\tau_3\over 2} q|0>}
wherein we sum over colour.
This can be written in terms of our proper vertex by reinstating the 
amputated legs.
One has from eq.\dfnfpi \ the exact result
\eqn\resfpi{if_{\pi} k^{\mu}=-(\gamma^{\mu}\gamma_5)^{\alpha \beta}
\sum_{\rm colour} \int d^4x 
\left[\left({1\over i\partial\hskip -.08in / _1}\right)
\left({1\over i\partial\hskip -.08in / _2}\right)
(i\partial\hskip -.08in /  _1)(i\partial\hskip -.08in /  _2)<\pi^0|u_{\beta}(x_1)
\bar{u}_{\alpha}(x_2)|0>\right]_{X=0}}
corresponding to 
the coordinate space representation of the diagram in Fig. 4.
Thus finally
\eqn\finfpi{if_{\pi} k^{\mu}=-3\int d^4x {\rm Tr} \left[\gamma^{\mu}\gamma_5 
\left({1\over i\partial\hskip -.08in / _1}\right)
\left({1\over i\partial\hskip -.08in / _2}\right)\hat{\Gamma}(x_1,x_2)\right]_{X=0}.}
This result does not rely on perturbation theory.
The free massless quark propagators $(1/ i\partial\hskip -.08in / )_{1,2}$ appear 
in equation \finfpi \ only because of our definition of $\Gamma .$

For the $\pi^0 \to\gamma^* +\gamma^*$ amplitude we shall single out 
the diagram depicted in Fig. 5.
Unlike the case of the calculation of $f_{\pi}$ the claim that this 
diagram is the most important one is by no means trivial.
As above, the quark propagators carrying the momenta $l\pm k/2$
appear as a result of our definition of $\Gamma (k,l)$,
so that here the use of free quark propagators does not necessarily
imply a perturbative approach.
The free propagator corresponding to momentum $l-q$ is meaningful
if the quark line has a large virtuality, $(l-q)^2\to \infty$,
which will be the case in our calculation.
Other diagrams, for example Fig. 6 with an additional gluon,
should be less important.
The reason is that if $q$ is a large momentum (or, more accurately, 
$q^2$ goes to infinity) then the biggest contribution arises when the 
flow of this momentum takes place through the minimal 
number of propagators.
In fact if in Fig. 6 the upper 4-point proper Green function
is a rapidly decreasing function of the virtualities of the external
particles, the large momentum $q$ will flow through the two quark
propagators at the bottom of the diagram.
The overall behaviour of the diagram will be $\sim 1/q^3$ or 
$\sim 1/q^4$, instead of $\sim 1/q^2$ for the diagram in Fig. 5.
In other words a soft hadronic system (the pion) will not tolerate
a large momentum transfer because of the rapid fall of the form factors
with increasing momentum transfer.

Another class of diagrams are the radiative corrections to
Fig. 5 such as shown in Fig.7.
In this case the flow of large momentum $q$ does not lead to any 
additional power decrease in $q^2$ because of the divergence of
the internal triangle diagram.
The vertex corrections replace the QCD coupling by the running 
coupling evaluated at $\approx q^2 .$
Due to asymptotic freedom this produces only logarithmic corrections to 
the diagram of Fig. 5.

It should be stressed that the diagram of Fig. 5 can be only 
relied upon for large momentum $q^2$. This is obvious when there 
is just one large space-time scale determined by $q_1^2=q_2^2=
-q^2=Q^2 .$
In this case the characteristic time interval between the emissions of 
the two photons is short, $\approx 1/Q^2$, and no additional 
interactions of the quark can take place.
This is certainly not true for small $q^2$.

\newsec{Calculation of $f_{\pi}$ and asymptotics at large $q_1^2=q_2^2$}

Passing to the momentum representation in eq.\finfpi \ and substituting 
the general structure for $\Gamma (k,l)$ (eq.\gener ) one obtains 
the following expression for $f_{\pi}$
\eqn\formfpi{f_{\pi} k^{\mu}= 12 \int{d^4l\over i(2\pi)^4}
\left\{\left[B +{1\over 2} (k\cdot l) C\right]{(l-k/2)^{\mu} \over (l-k/2)^2} -
\left[B -{1\over 2} (k\cdot l) C\right]{(l+k/2)^{\mu} \over (l+k/2)^2}\right\} .}
The form factors $A$ and $D$ disappear when the trace is calculated.
We change $l\to -l$ in the second integral, use the fact that $B$ and 
$C$ are even functions of $l$, and can substitute 
$l_{\mu}\to k_{\mu} (k\cdot l)/k^2$ under the integral sign to obtain
\eqn\ffpi{f_{\pi}= -12 \int {d^4l\over i(2\pi)^4}\left[B +{1\over 2} (k\cdot l) C\right]{1 
\over (l-k/2)^2}\left(1-2{(k\cdot l)\over k^2}\right) .}
For the amplitude $\gamma^* (q_1)\to \gamma^* (q_2)+\pi^0 (k)$ the direct 
calculation of the diagram of Fig. 5 yields the result (leaving out the
polarization vectors) 
$$ \;$$
\eqn\pigg{
{\cal M}_{\mu\nu}(q,k)=4e^2\epsilon_{\mu\nu\lambda\sigma}\int 
{d^4l\over i(2\pi)^4}
{1\over (l-q)^2} \left\{ {q^{\lambda}k^{\sigma}-l^{\lambda}
k^{\sigma}-2q^{\lambda}l^{\sigma}\over (l-k/2)^2}\left[B+{1\over 2}(k\cdot l)C\right] +\right.}
$$\left.+{q^{\lambda}k^{\sigma}-l^{\lambda}
k^{\sigma}+2q^{\lambda}l^{\sigma}\over (l+k/2)^2}\left[B-{1\over 2}(k\cdot l)C\right] \right\}.$$
By changing $l\to -l$ and $q\to -q$ in the second integral we can represent 
the amplitude ${\cal M}$ in the form
\eqn\piggy{{\cal M}_{\mu\nu}(q,k)=4e^2\epsilon_{\mu\nu\lambda\sigma}
[A^{\lambda\sigma}(q,k)-A^{\lambda\sigma}(-q,k)],}
where 
$$A^{\lambda\sigma}(q,k)=\int {d^4l\over i(2\pi)^4}
{1\over (l-q)^2} {q^{\lambda}k^{\sigma}-l^{\lambda}
k^{\sigma}-2q^{\lambda}l^{\sigma}\over (l-k/2)^2}
\left[B+{1\over 2}(k\cdot l)C\right] .$$
The integrals of $l^{\sigma}$ multiplied by scalar factors must yield 
results proportional to $q^{\sigma}$ or $k^{\sigma}$.
This is tantamount to changing 
\eqn\ch{l^{\sigma}\to q^{\sigma}{(l\cdot k)k^2-
(l\cdot k)(k\cdot q) \over k^2q^2-(k\cdot q)^2}+
k^{\sigma}{(l\cdot k)q^2-(l\cdot q)(k\cdot q) \over k^2q^2-(k\cdot q)^2} }
under the integral sign.

The final result is given by 
\eqn\fpig{{\cal
M}_{\mu\nu}=4e^2\epsilon_{\mu\nu\lambda\sigma}q^{\lambda}k^{\sigma}
[A(k,q)+A(k,-q)],}
$$A(q,k)=\int {d^4l\over i(2\pi)^4}
{(l-q)\cdot V(q) \over (l-q)^2 (l-k/2)^2}[B+(k\cdot l)C/2] ,$$
with 
\eqn\nothing{V^{\mu} (q)={1\over k^2q^2-(k\cdot q)^2}[k^{\mu}(k\cdot q -2q^2)+
q^{\mu}(2 k\cdot q-k^2)] .}
One can show that for $q_1^2=q_2^2 \approx q^2$ and 
$|q^2|\to\infty$ the term $(l-q)^2$ can be replaced by $q^2 .$
We shall analyze below in detail when it is safe to make this 
replacement.
For the moment let us accept it and notice that eq.\fpig \ then involves
$$[V^{\mu} (q)-V^{\mu} (-q)]\cdot q^{\mu}=-2,$$
and
$$[V^{\mu} (q)+V^{\mu} (-q)]l^{\mu}={4\over k^2q^2-(k\cdot q)^2}[(q\cdot l)(q\cdot k)-
q^2(k\cdot l)],$$
and the remaining terms in the integral do not depend on $q$, so that 
effectively $l^{\lambda}\to k^{\lambda}(k\cdot l)/k^2$ under the integral sign.

Taking all this into account we readily get
\eqn\tired{
{\cal M}_{\mu\nu}(q,k)=8e^2\epsilon_{\mu\nu\lambda\sigma}q^{\lambda}
k^{\sigma} {1\over q^2} \int {d^4l\over i(2\pi)^4}{1\over (l-k/2)^2}
\left[B+{1\over 2}(k\cdot l)C\right] \left(1-2{(k\cdot l)\over k^2}\right).}
We see that up to a numerical factor the last integral 
is nothing other than $f_{\pi}$ in eq.\ffpi .
Comparing with eq.(1.1) we see that 
\eqn\M{M(q_1^2,q_2^2)=-{2e^2\over 3}\cdot {f_{\pi}\over q^2}}
for $q_1^2=q_2^2=q^2$ and $|q^2|\to \infty ,$
which coincides with the OPE result \volo \ for $q^2\to -\infty$
given in eqs.(1.3) and (1.4).

\newsec{Asymptotics for $q_1^2\neq q_2^2$}

To analyse the situation at $q_1^2\neq q_2^2$
we choose the pion rest frame as a reference frame.
If we also choose the $z$ axis to be directed along $\vec{q}=(\vec{q}_1
+\vec{q}_2)/2$, one has for the time and third components of $k^{\mu}$ and 
$q^{\mu}$
\eqn\frame{k^{\mu}=(k_0,k_3)=(m_{\pi},0),
\;\; q^{\mu}=(q_0,q_3)=(k\cdot q/m_{\pi},\sqrt{(k\cdot q)^2-k^2 q^2}/m_{\pi}),}
where 
\eqn\inv{q^2={1\over 2} (q_1^2+q_2^2)-{1\over 4}m_{\pi}^2
\approx {1\over 2} (q_1^2+q_2^2), \;\; (k\cdot q)={1\over 2} (q_1^2-q_2^2).}
We are interested in the limit $|q_1^2|,|q_2^2|\to \infty$ 
at fixed $\omega$, where
\eqn\omdef{\omega={(k\cdot q)\over q^2}=
{q_1^2-q_2^2\over q_1^2+q_2^2},\;\;
-1\leq \omega\leq 1.}
The condition $|\omega|\leq 1$ follows from our assumption
that $q_1^2$ and $q_2^2$ are of the same sign which is the case of physical 
interest.
In this limit one has for $k^{\mu}$ and $q^{\mu}$ 
\eqn\kq{k^{\mu}=m_{\pi} (1,0,0,0),\;\; 
q^{\mu}={(k\cdot q)\over m_{\pi}}(1,0,0,1).}
One sees that, since in our reference frame the components
of $l_{\mu}$ are of the order of hadronic scale ($\sim 1 \; GeV$),
the scalar product $(l\cdot q)$ is of the order of $(k\cdot q)$, or more 
accurately 
\eqn\lqu{(l\cdot q)={(k\cdot q)\over m_{\pi}}(l_0-l_3).}
Using this we can simplify the general expression \fpig \ at $(k\cdot q)^2>>
k^2 q^2$ ($k^2=m_{\pi}^2$) and obtain
\eqn\ampone{{\cal M}=
4e^2\epsilon_{\mu\nu\lambda\sigma} \epsilon^{\mu}(q_1)
\epsilon^{\nu} (-q_2) q^{\lambda}k^{\sigma} \times }
$$\int {d^4l\over i(2\pi)^4}
{B+(k\cdot l)C/2\over (l-k/2)^2}\left[{1\over (l-q)^2}+{1\over (l+q)^2}\right]
\left(1-{2(l\cdot q) \over (k\cdot q)}\right).$$
In our reference frame eq. \ampone \ can be rewritten in the form
\eqn\amptwo{{\cal M}={4e^2\epsilon_{\mu\nu\lambda\sigma}  
\epsilon^{\mu}(q_1) \epsilon^{\nu} (-q_2) q^{\lambda}k^{\sigma}
\over q^2}   \times}
$$\times \int {d^4l\over i(2\pi)^4}
{B+m_{\pi}l_0
C/2\over l^2-m_{\pi} l_0}\left[{1\over 1-2(l_0-l_3)\omega/m_{\pi}}
+{1\over 1+2(l_0-l_3)\omega/m_{\pi}}\right]\times$$
$$\times \left(1-{2(l_0-l_3)\over m_{\pi}}\right).$$
In the last expression we neglected $l^2/q^2<<1$ in the denominators
in the square brackets and $m_{\pi}^2/4$ in the first denominator.

Consider now the limit $m_{\pi}\to 0$ (the chiral limit).
At first glance the amplitude ${\cal M}$ has a natural 
dependence on the parameter $\omega/m_{\pi}$ and one might expect 
a rapid variation of the function ${\cal M}$ at $\omega\sim m_{\pi}/\Lambda
<< 1$, where $\Lambda\sim 1 \; GeV$ is a characteristic hadronic scale.
This is not, however, correct.
 
Expanding the denominators in the square brackets in eq. \amptwo \ we get a series 
in the parameter $[\omega (l_0-l_3)/m_{\pi}]^2$.
It may seem that the terms of this series are singular at $m_{\pi}\to 0.$
However the integrals multiplying all negative powers of $m_{\pi}$
vanish.

To see this let us expand all the other factors i.e. the denominators 
$(l^2-m_{\pi} l_0)^{-1}$and the functions $B$ and $C$ which depend 
on $(k\cdot l)=m_{\pi} l_0$, in the parameter $(m_{\pi} l_0)$.
The integral  \amptwo \  becomes
$$\;$$
\eqn\Mexp{{\cal M}= 8e^2 \epsilon_{\mu\nu\lambda\sigma} 
\epsilon^{\mu}(q_1) \epsilon^{\nu} (-q_2)
q^{\lambda}k^{\sigma}\times }
$$\times \int {d^4 l\over i(2\pi)^4}
\left[\sum^{\infty}_{k=0} \omega^{2k}\left({2(l_0-l_3)\over m_{\pi}}\right)^{2k}-
\sum^{\infty}_{k=0} \omega^{2k}\left({2(l_0-l_3)\over m_{\pi}}\right)^{2k+1}\right]\times$$
$$\times\sum^{\infty}_{p=0}(m_{\pi}l_0)^p A_p (l^2)$$
where we have introduced the expansion
\eqn\anexp{{B+m_{\pi}l_0 C/2\over l^2-m_{\pi}l_0}=
\sum^{\infty}_{p=0}A_p(l^2)(m_{\pi}l_0)^p.}
We do not specify at the moment the functions $A_p(l^2)$.
In the expression \Mexp \ one can rotate $l_0\to i l_4$  since the
singularities of $A_p(l^2)$ in $l_0$ obey the Feynmann rules;
their positions in the complex $l_0$ plane are determined by
$l_0=\pm \sqrt{\Lambda^2+\vec{l}^2}-i\epsilon$ where $\Lambda$ is a mass
parameter.
This leads to the integration in euclidean space
\eqn\meas{{d^4l\over i(2\pi)^4}\to  {d^4l_E\over (2\pi)^4},\;\;
d^4l_E=dl_3 dl_4 d^2l_{\perp},\;\; d^2l_{\perp}=dl_1 dl_2.}
We now introduce 
\eqn\morenotations{l_4=l_{||} \cos \phi,\;\; l_3=l_{||}\sin \phi}
and see that the nonvanishing contributions in the r.h.s. of eq. \Mexp \
come only from the terms with $p\geq 2k$ in the first sum over $p$
and with $p\geq 2k+1$ in the second one.
Indeed
\eqn\moremore{l_0-l_3=il_4-l_3=il_{||}e^{i\phi},\;\;
l_0={i\over 2}l_{||}(e^{i\phi}+e^{-i\phi}),}
and the integration over $\phi$ yields zero contribution unless 
the power of $l_0$ is larger than or equal to that of $(l_0-l_3).$

Therefore the amplitude can be represented as a double series in
$\omega^2$ and $m_{\pi}^2 .$
We shall now try to calculate the leading term in $m_{\pi}^2$ for which 
purpose one should leave in eq. \Mexp \ only the contributions 
from $p=2k$ and $p=2k+1$.

One has 
\eqn\Mthree{{\cal M}={2e^2\epsilon_{\mu\nu\lambda\sigma} 
\epsilon^{\mu}(q_1) \epsilon^{\nu} (-q_2)
q^{\lambda}k^{\sigma} \over 4\pi^2q^2}\times}
$$\times \int_0^{\infty}dl_E^2\int_0^{l_E^2}dl_{||}^2
\left[\sum_{k=0}^{\infty}\omega^{2k}(l_{||}^2)^{2k} A_{2k}+
\sum_{k=0}^{\infty}\omega^{2k}(l_{||}^2)^{2k+1} A_{2k+1}\right].$$
The integrals over $l_{||}^2$ are readily calculated and one obtains
\eqn\Mfour{{\cal M}={2e^2\epsilon_{\mu\nu\lambda\sigma} 
\epsilon^{\mu}(q_1)
\epsilon^{\nu} (-q_2)
q^{\lambda}k^{\sigma} \over 4\pi^2q^2}
\int_0^{\infty}dl_E^2\sum_{k=0}^{\infty} \left({(l_E^2)^{2k+1} A_{2k}
\over 2k+1}+ {(l_E^2)^{2k+2} A_{2k+1}
\over 2k+2}\right) \omega^{2k}.}
To determine the coefficients $A_{2k}$ and $A_{2k+1}$
we expand the factors entering eq. \anexp \ 
\eqn\BCmore{{B+m_{\pi}l_0 C/2\over l^2-m_{\pi}l_0}=
\sum_{p=0}^{\infty} (B_p+{1\over 2}m_{\pi} l_0 C_p)(m_{\pi}l_0)^{2p}
\sum_{q=0}^{\infty}{m_{\pi}^q l_0^q\over (l^2)^{q+1}},}
where we have used the fact that $B$ and $C$ are the even functions
of $(k\cdot l)=m_{\pi}l_0.$
The coefficients $A_{2k}$ and $A_{2k+1}$ in \Mfour \ are equal to
\eqn\Ak{A_{2k}=\sum^k_{p=0}{B_p\over (l^2)^{2k-2p+1}}+
{1\over 2}\sum_{p=0}^{k-1}{C_p\over (l^2)^{2k-2p}},}
$$A_{2k+1}=\sum^k_{p=0}{B_p\over (l^2)^{2k-2p+2}}+
{1\over 2}\sum_{p=0}^{k}{C_p\over (l^2)^{2k-2p+1}}.$$
It is implied that in the expression \Ak \ for $A_{2k}$ the terms proportional
to $C_p$ are absent for $k=0$.

Substituting \Ak \ into \Mfour \ and changing the order of summation
in $p$ and $k$ one can derive the following expression 
for ${\cal M}$
\eqn\Mfive{{\cal M}= -{2e^2\epsilon_{\mu\nu\lambda\sigma} \epsilon^{\mu}(q_1)
\epsilon^{\nu} (-q_2)
q^{\lambda}k^{\sigma}\over 
8\pi^2q^2} \times}
$$\times \sum^{\infty}_{p=0} \left\{\int_0^{\infty}  dl_E^2\;
\left[B_p (l_E^2) +{1\over 2}l_E^2C_p(l_E^2)\right]
(l_E^2)^{2p} \phi_p(\omega)+\right.$$
$$\left. +\int_0^{\infty}  dl_E^2\;
(l_E^2)^{2p+1}C_p(l_E^2)\left({\omega^{2p} \over 2p+1}-\phi_p(\omega)\right) \right\},$$
where 
\eqn\phip{\phi_p(\omega)=\sum_{k=p}^{\infty}{\omega^{2k}\over (k+1)(2k+1)},}
$$ \phi_0 (\omega)=\phi (\omega) ={1\over \omega^2}[(1+\omega)
{\rm ln}(1+\omega)+(1-\omega){\rm ln}(1-\omega)].$$
Generally the coefficient functions $B_p(l_E^2)$ and $C_p(l_E^2)$
in the expansion of $B(l^2,(k\cdot l)^2)$ and $C(l^2,(k\cdot l)^2)$ in $(k\cdot l)^2$
are unknown.
However we shall see that if the form factors $B$ and $C$ decrease fast 
enough at infinity as functions of the quark masses (virtualities)
$m_1^2=(l+k/2)^2$ and $m_2^2=(l-k/2)^2$, then the coefficients
$B_p$ and $C_p$ decrease with $p$ and since also the $\phi_p (\omega)$
decrease with $p$ one can obtain a reasonable estimate by keeping only the first
term in the sum in eq. \Mfive .

To see this we consider a simple model when the mass dependence 
of $B$ (or $C$) is determined by
\eqn\Bmod{B=\left({1\over (\Lambda^2-m_1^2)(\Lambda^2-m_2^2)}\right)^n}
where we omit an inessential constant factor; and $n\geq 1$ is an integer.

For this model the coefficients $B_p(l_E^2)$ read
\eqn\Bpmod{B_p={n(n+1)\cdot\cdot\cdot (n+p-1)\over p!}\cdot {1\over (\Lambda^2+l_E^2)^{2n+2p}},\;\; p\geq1,}
$$B_0={1\over (\Lambda^2+l_E^2)^{2n}}.$$
The integrals in eq. \Mfive \ are of the form 
\eqn\Ipmod{I_p=(\Lambda^2)^{2n-1}
\int^{\infty}_0 dl_E^2 \; l_E^{4p} B_p(l_E^2)=}
$$={n(n+1)\cdot\cdot\cdot (n+p-1)\over p!}\cdot
{(2p)!\over (2n-1)2n(2n+1)\cdot\cdot\cdot (2n+2p-1)}.$$
Therefore 
\eqn\III{I_0={1\over 2n-1},\;\; I_1={1\over (2n-1)(2n+1)},\;\; 
I_2={3\over (2n-1)(2n+1)(2n+3)},}
$$ I_3={15\over (2n-1)(2n+1)(2n+3)(2n+5)},\;\; ...$$
We see that the ratio $I_1/I_0=1/(2n+1)$ is only $30\%$ for $n=1$ and  $20\%$
for $n=2$, while the other coefficients are even smaller.

Keeping only the $p=0$ terms in eq. \Mfive \ one obtains 
\eqn\Msix{{\cal M}=-{2e^2\epsilon_{\mu\nu\lambda\sigma} \epsilon^{\mu}(q_1)
\epsilon^{\nu} (-q_2)
q^{\lambda}k^{\sigma}\over q^2}
[f_{\pi}\phi(\omega)/3+g_{\pi} (1-\phi(\omega))],}
where
\eqn\mmm{g_{\pi}={1\over 8\pi^2}\int^{\infty}_0 dl_E^2 l_E^2 C_0(l_E^2)}
and we have used the fact that the expression \ffpi \ for $f_{\pi}$ simplifies 
in the chiral limit to
\eqn\fpisimp{f_{\pi}={3\over 8\pi^2} \int_0^{\infty}dl_E^2 \left(B_0(l_E^2)+
{1\over 2}l_E^2 C_0(l_E^2)\right).}
Eq. \Msix \  provides an expression for the amplitude ${\cal M}$ in terms 
of one unknown parameter $g_{\pi}$ and, as such, is of some interest.
We shall argue however that $g_{\pi}<<f_{\pi}$ and thus the term 
proportional to $g_{\pi}$ may be neglected.
In the next section we show that $B(0)=1/f_{\pi}$.
Thus $B(0)$ describes the low energy interaction of pions with quarks
(see eq. (1.9)) reflecting the Goldstone character of pions.
The form factor $C(0)$ does not show up at all in this interaction
at low energies if the PCAC ideology is used.
Since $f_{\pi}=93~ MeV$ is ``abnormally'' small in the units of usual hadronic 
scale $\Lambda\sim 1~GeV$,
the value of  $B(0)$ is ``abnormally'' large, but we would expect that 
$C(0)$ has a ``natural'' value $C(0)\sim 1/\Lambda^3$.
If this is correct then
the term involving $C$ in eq. \fpisimp \ should be negligible and we would 
expect roughly
\eqn\fpimore{f_{\pi}\sim {\Lambda^2 B(0)\over 8\pi^2}=
{\Lambda^2\over 8\pi^2f_{\pi}}.}
Then for $g_{\pi}$ in \mmm \ we expect 
\eqn\gppp{g_{\pi}=\sim {\Lambda^4 C(0)\over 8\pi^2}
\approx {\Lambda\over  8\pi^2} = {f_{\pi}\over \Lambda}\cdot f_{\pi}}
so that $f_{\pi}\approx10g_{\pi} .$

Neglecting the $g_{\pi}$-term in eq. \Msix \  we arrive at the final result
\eqn\Mseven{{\cal M}= -{2e^2\epsilon_{\mu\nu\lambda\sigma} 
\epsilon^{\mu}(q_1)
\epsilon^{\nu} (-q_2)
q^{\lambda}k^{\sigma}\over 3q^2}
f_{\pi}\phi(\omega) ,}
$$\phi(\omega)={1\over \omega^2}[(1+\omega){\rm ln} (1+\omega)+
(1-\omega){\rm ln} (1-\omega)]= \sum^{\infty}_{k=0}{\omega^{2k}\over 
(k+1)(2k+1)}$$
which was quoted in Introduction.

In getting eq. \Mseven \ we neglected the second term $\sim g_{\pi}$ and 
all terms with $p\neq 0$ of eq. \Mfive .
Since for $n=2$, which seems a reasonable 
value in the model (4.19), the accuracy of the second of these approximations 
is about $20\%$ and there is an additional smallness related to the fact that 
$\phi_{p+1}(1)
{\ \lower-1.2pt\vbox{\hbox{\rlap{$<$}\lower5pt\vbox{\hbox{$\sim$}}}}\ }
0.3 \phi_p(1) $ we expect the accuracy of the final result \Mseven \
might be of order 
${\ \lower-1.2pt\vbox{\hbox{\rlap{$<$}\lower5pt\vbox{\hbox{$\sim$}}}}\ }
10\%$.
The value $\phi(0)=1$ corresponds to the OPE result \volo \ for the asymptotics 
at $q_1^2=q_2^2$.
On the other hand, $\phi(1)=2\;{\rm ln}\; 2\approx 1.4$.
This has to be compared with the result $\phi(1)=3/2$ given in ref. \marc \ 
(see Introduction). 
Since we used certain numerical approximations and our whole approach 
is different from that of ref. \marc \  the $7\%$ coincidence of these two figures
seems to be quite satisfactory.
We plot this function $\phi (\omega)$ in Fig. 10.
It is interesting to compare our result for the function $\phi (\omega)$ with that 
given in ref. \brod .
In notation of ref. \gorsky \ the function $\phi$ (denoted there as $I(\omega)$)
is represented by an integral 
$$I(\omega)=\int^1_{-1} {d\xi \phi_A (\xi,\mu^2)\over 1-\omega\xi} .$$
In the limit of infinitely large renormalization scale $\mu\to \infty$
we have
\rad\ \zh\
\brod\  \ref\CZ{V.L. Chernyak and A.R. Zhitnitsky, Phys. Rep. {\bf 112} (1984) 173.}
$$\phi_A (\xi,\mu^2)\to \phi_A^{as} (\xi)={3\over 4} (1-\xi^2).$$
As it is pointed out in ref. \gorsky \ this expression for $\phi_A (\xi,\mu^2)$ is 
reliable only for $|\omega|<1/2$.
To illustrate the difference between the functions $\phi$ and $I$ 
we plotted the ratio $2( \phi -I)/(\phi+I)$ in Fig. 11.
It is easy to see that for $|\omega|<1/2$ the difference is~1\% .
At $|\omega|=1$ the difference is 8\% which is not surprising since we use quite different 
physical assumptions.
Thus the above estimate of our accuracy is compatible with 
the difference of our result with that of ref. \brod .

\newsec{$\pi^0\to 2\gamma$ and the anomaly}

It is interesting to understand how the formalism developed in this paper corresponds to the description of $\pi^0$ decay into real photons by PCAC with the Adler-Bell-Jackiw
anomalous term.
We shall see below that the comparison of these two approaches 
leads to the relation 
\eqn\Bfpi{B(0)=1/f_{\pi} ,}
where $B(0)$ is the value of the form factor $B$ with all momenta zero.

First of all it should be stressed that since $q^2=(q_1+q_2)^2/4=-m_{\pi}^2/4$
is now small there is no reason to consider only the triangle diagram
which has been used above.
The most general amplitude of $\pi^0\to 2\gamma$ decay can be represented as shown 
in Fig. 8.
The ${\cal M}_2$ amplitude of Fig. 8$b$ is of the most general form provided 
that the four-point Green function $v_{\nu}$ is of the most general character.
In the ${\cal M}_2$ amplitude we have arbitrary chosen to show explicitly the quark
which emits the $(q_1,\mu)$ photon while the vertex for the emission 
of the second photon, $(q_2,\nu)$, is not shown and is absorbed into $v_{\nu}$.
Separating the ${\cal M}_1$ amplitude (Fig. 8$a$) from ${\cal M}_2$ 
means that we exclude  the pole 
contribution to $v_{\nu}$ depicted in Fig. 9. 
The reason for doing so is that in this section we shall intensively use the chiral limit,
together with Adler's condition for the emission of soft pions.
This condition requires that any amplitude should vanish in the limit when the pion 
four-momentum goes to zero except the pole terms which can survive.
Because of Adler's condition $v_{\nu}$ being defined without 
a pole contribution
vanishes at $k_{\mu}\to 0$, i.e. we have
\eqn\Vnu{v_{\nu}\to 0 \;\; {\rm at}\;\; k_{\mu}\to 0 .}
The idea of the argument below is to prove that in computing the amplitude 
${\cal M}_2$ we can use the Green function $v_{\nu}$ at $q_2=0$.
Given this we can use the Ward identity for the emission of 
a photon with zero momentum
\eqn\Vward{v_{\nu} (k,l,0)=-{\partial \Gamma (k,l)\over \partial l^{\nu} }.}
Thus we shall manage to compute ${\cal M}_2$ in terms of the $\Gamma (k,l)$
vertex used throughout this paper.

We now give to the proof that  $v_{\nu} (k,l,q_2)$ can indeed be taken 
at $q_2=0$.

The most general structure of  $v_{\nu}$ which survives in the calculation of the 
amplitude ${\cal M}_2$ is 
\eqn\Vnumore{v_{\nu} (k,l,q_2)=
(F_{1\nu}k\hskip -.08in / +F_{2\nu}l\hskip -.07in / +F_{3\nu}q\hskip -.07in / _2 +
F_{4}\gamma_{\nu})\gamma_5 ,}
where all functions $F_{1\nu},$ $F_{2\nu}$, $F_{3\nu}$ and $F_{4}$ depend on the  
vectors $k,$ $l,$ and $q_2.$

Calculating the trace appearing in computation of the amplitude 
${\cal M}_2$ (Fig. 8$b$) 
one obtains the following expression
\eqn\tc{\epsilon^{\lambda\rho\mu\sigma}
[F_{1\nu}l_{\rho}q_{2\sigma}k_{\lambda} +
F_{2\nu}k_{\rho}q_{2\sigma}l_{\lambda}/2+
F_{3\nu}l_{\rho}k_{\sigma}q_{2\lambda}+
F_{4}(l_{\rho}k_{\sigma}-l_{\rho}q_{2\sigma}+k_{\rho}q_{2\sigma}/2)] .}
Consider each of the terms in \tc .

For the $F_{1\nu}$ term we already have the necessary structure to provide 
the $\epsilon_{\mu\nu\lambda\sigma} \epsilon^{\mu}(q_1)$
$\epsilon^{\nu} (-q_2)
q^{\lambda}k^{\sigma}$ factor
(since $\epsilon_{\lambda\rho\mu\sigma}q_{2\sigma}k_{\lambda}=
\epsilon_{\lambda\rho\mu\sigma}q_{2\sigma}q_{1\lambda}$).
Therefore the contribution proportional to $F_{1\nu}$ is already of the second 
order in the small momenta (photon momenta) and thus all the other 
factors can be taken at zero momenta. 
Particularly we can take $F_{1\nu}$ at $q_2=0$.

For the $F_{2\nu}$ term the situation is basically the same and again $F_{2\nu}$
can be taken at $q_2=0$.
In this case, however, $F_{2\nu}$ does not actually contribute since in 
Adler's soft pion limit $F_{2\nu}\to 0$ as $k_{\mu}\to 0.$
This follows from an application of Adler's soft pion theorem 
to $F_{2\nu}(k,l,q_2)$.

The $F_{3\nu}$ form factor can also be taken at $q_2=0$ but now this is not the same 
as putting $q_2=0$ in $v_{\nu}$ (unlike the case of the first two terms in \Vnumore )
since $F_{3\nu} (k,l,0)q\hskip -.07in / _2$ in principle could survive.
Luckily the whole contribution to ${\cal M}_2$, proportional to $F_{3\nu}$, is again 
of the third order in the small momenta.
Indeed $k_{\sigma}q_{2\lambda}\to q_{1\sigma}q_{2\lambda}$ and also $F_{3\nu}$
is itself proportional to the pion 4-momentum due to the soft pion emission 
theorem.

Lastly the $F_4$ form factor can also be taken at $q_2=0$.
Indeed part of it, proportional to $(k\cdot q_2)$ is immediately negligible, while 
the part  $\propto k\cdot l$ can contribute through $F_4l_{\rho}q_{2\sigma}\to 
F_4 q_{1\rho}q_{2\sigma}$, but then $F_4$ can obviously be taken at $q_2=0$.

Thus we see that for different reasons all the 
contributions to $v_{\nu}$ in eq. \Vnumore \ can be taken at $q_2=0$.
Therefore $v_{\nu}(k,l,q_2)\to v_{\nu}(k,l,0)$ and the fundamental Ward 
identity \Vward \ is applicable to our calculation.

Using for $\Gamma (k,l)$ the expression 
\eqn\Gexp{\Gamma (k,l)=[iBk\hskip -.08in / +i(k\cdot l)C l\hskip -.07in / ]\gamma_5 }
and differentiating it w.r.t. $l_{\nu}$ we need to keep only the terms 
\eqn\Vnutwo{v_{\nu}=-i[2k\hskip -.08in / l_{\nu}{dB(l^2)\over dl^2} +
(k\cdot l)C\gamma_{\nu}]\gamma_5 }
as follows from the consideration given above: only the $F_{1\nu}$ and $F_4$ 
form factors can actually contribute to the amplitude in question.

We are now ready to write down the final result for ${\cal M}_1$ and 
${\cal M}_2$ 
of Fig. 8 which can be obtained by a straightforward calculation.
For ${\cal M}_2$ we obtain
\eqn\Ttwo{{\cal M}_2=
{2e^2\epsilon_{\mu\nu\lambda\sigma} \epsilon^{\mu}(q_1)
\epsilon^{\nu} (-q_2)
q^{\lambda}k^{\sigma}\over 8\pi^2} \left[B(0)-{1\over 2}\int^{\infty}_0
dl^2_E C(l_E^2)\right].}
For the amplitude ${\cal M}_1$ the result is 
\eqn\Tone{{\cal M}_1= {2e^2\epsilon_{\mu\nu\lambda\sigma} \epsilon^{\mu}(q_1)
\epsilon^{\nu} (-q_2)
q^{\lambda}k^{\sigma}\over 16\pi^2} \int^{\infty}_0
dl^2_E C(l_E^2) .}
Thus the sum of ${\cal M}_1$ and ${\cal M}_2$ 
does not contain $C$ and there results 
\eqn\Tsum{{\cal M}={\cal M}_1+{\cal M}_2={2e^2\epsilon_{\mu\nu\lambda\sigma} \epsilon^{\mu}(q_1)
\epsilon^{\nu} (-q_2)
q^{\lambda}k^{\sigma}\over 8\pi^2}B(0).}
We can compare this result with the usual PCAC amplitude
\eqn\Tpcac{{\cal M}={2e^2\epsilon_{\mu\nu\lambda\sigma} \epsilon^{\mu}(q_1)
\epsilon^{\nu} (-q_2)
q^{\lambda}k^{\sigma}\over 8\pi^2 f_{\pi}}}
to get the relation 
\eqn\relB{B(0)={1\over f_{\pi}}.}
This equation can be understood in the following way.
In the chiral limit of vanishing 4-momentum of the pion the pion-quark-antiquark
interaction has a pseudovector character (corresponding to the
$B$-form factor) with a fixed coupling constant
\eqn\Leffmore{L_{eff}={1\over f_{\pi}} (\partial_{\mu}\Phi_{\pi})
[\bar{u}\gamma_{\mu}\gamma_5  u-\bar{d} \gamma_{\mu}\gamma_5 d].}
Equation \Leffmore \ reveals the Goldstone character of the pion.

\newsec{Conclusions}

We have studied the $\pi^0 \gamma\gamma$ vertex for a range of virtual 
photon squared masses $q_1^2, q_2^2$, in a rather general picture  where
the pion is treated as a composite, non-pointlike particle.
We show how our approach reproduces the operator product expansion 
results valid for $q_1^2=q_2^2\to \infty$ and compare with the infinite momentum frame 
result for $q_2^2=0$, $q_1^2\to \infty .$

We have been able to find a closed analytic formula valid for arbitrary 
fixed values of the ratio $q_1^2/q_2^2$ in the limit that at least one of 
$|q_1^2|$ and $|q_2^2|$ is large.
If the Feynman amplitude is written 
$${\cal M}=M(q_1^2, q_2^2) \epsilon_{\mu\nu\lambda\sigma}\epsilon^{\mu} (q_1)
\epsilon^{\nu}(-q_2) q^{\lambda}k^{\sigma}$$
where $q=(q_1+q_2)/2$, $k=q_1-q_2$ and if we define 
$$\omega={q_1^2-q_2^2\over q_1^2+ q_2^2}, \;\;\;
-1\leq \omega\leq 1$$
then our result is 
$$M(q_1^2, q_2^2)= 
M(0,0) \cdot \left(-{8\pi^2 f_{\pi}^2 \over 3q^2}  \right)
\phi (\omega)$$
with 
$$\phi (\omega)={1 \over \omega^2}
[(1+\omega)\log (1+\omega)+(1-\omega)\log (1-\omega)].$$
The amplitude ${\cal M}$ is of great interest theoretically because of its link to the axial 
anomaly.
Indeed this connection has led to dramatically  incorrect estimates 
of the amplitude at large photon virtualities.

On the experimental side the amplitude ${\cal M}$ plays a crucial r\^ole 
in several physical processes.
For $q_1^2<0$, $q_2^2<0$ it is relevant to the two-photon physics reaction 
$e^+ e^-\to \pi^0 e^+ e^-$, and for $q_1^2>0$, $q_2^2>0$ it controls the annihilation
process $e^+ e^-\to \pi^0 l^+ l^-$.
It would be of great interest to test our prediction experimantally.

\newsec{Acknowledgements}
The authors thank N. Dombey, A. Gorsky, A. Vainshtein and M. Voloshin for interesting 
and valuable discussions, and V.A. Radyushkin for useful communications.

The authors are also grateful to the Royal Society for generously supporting a Joint 
Project  between Birkbeck College, London, and the St. Petersburg Nuclear Physics 
Institute and for a Visiting Fellowship under the European Science Exchange 
Programm, which made possible the collaboration with the Soltan Institute 
for Nuclear Studies, Warsaw.

The work of \L. \L. 
is supported in part by Polish KBN Grant  2-P 302 -143-06.
The work of A.A. and A.J. is partially supported by INTAS grant  (1)93-283.
The work of A.J. is also supported in part by the
Packard Foundation and by NSF grant PHY-92-18167.

\listrefs

{\bf Figure captions}
\vskip .5in

Fig. 1. Feynman diagram for $e^- +e^- (e^+) \to e^- +e^- (e^+)+\pi^0$ reaction.

Fig. 2. Feynman diagram for $e^- +e^+ \to e^- +e^++\pi^0$ and 
$e^- +e^+ \to \mu^- +\mu^++\pi^0$ reactions.

Fig. 3. Proper vertex for $\pi^0\to q\bar{q}$ decay.

Fig. 4. Diagram for calculation of $f_{\pi}$.

Fig. 5. Diagram for calculation of the $\pi^0\to \gamma^*+\gamma^*$ amplitude.

Fig. 6. Possible gluon corrections to the diagram of Fig. 5.

Fig. 7. Possible radiative corrections to the diagram of Fig. 5.

Fig. 8. a), b) Diagrams for contributions to the 
$\pi^0\to 2\gamma$ amplitude.

Fig. 9. Diagram for the pole contribution to the amplitude.

Fig. 10. Plot of the function $\phi (\omega)$.

Fig. 11. Plot of the ratio $\Delta (\omega)=2(\phi (\omega)-I (\omega))/
(\phi (\omega)+I (\omega))$, where $I (\omega)$ 
is the result of ref. \gorsky .

\input epsf
%%Begin InstantTeX Picture
\let\picnaturalsize=N
\def\picsize{3.0in}
\def\picfilename{fpi1.eps}
%If you do not have the picture file add:
%\let\nopictures=Y
%to the beginning of the file.
\ifx\nopictures Y\else{\ifx\epsfloaded Y\else\fi
\global\let\epsfloaded=Y
\centerline{\ifx\picnaturalsize N\epsfxsize \picsize\fi
\epsfbox{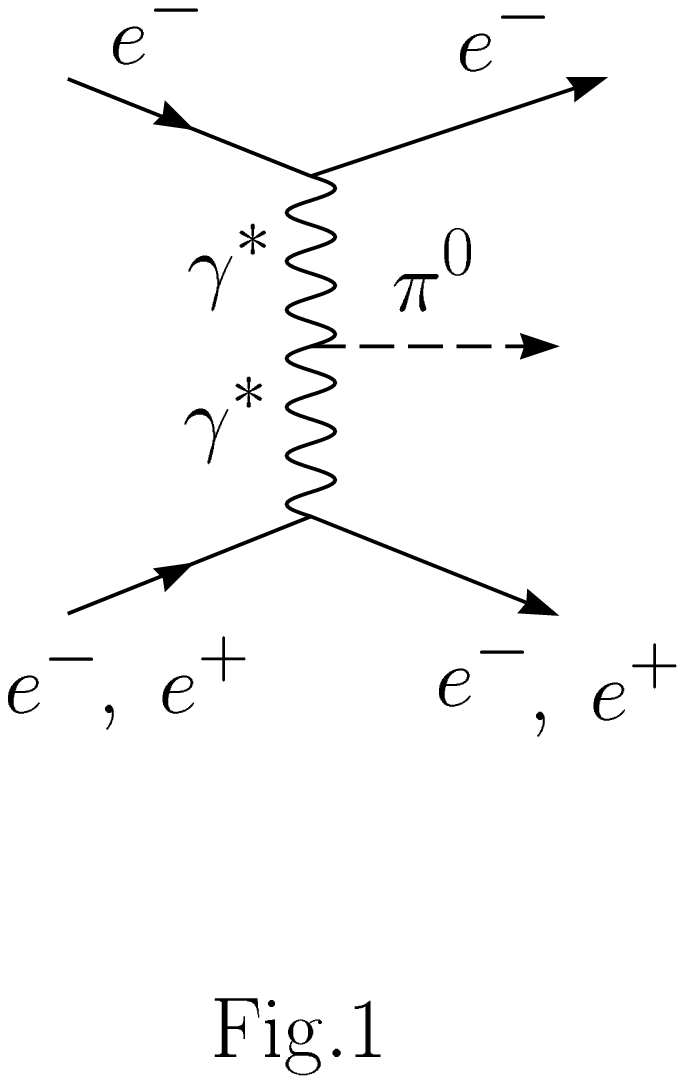}}}\fi
%%End InstantTeX Picture

%%Begin InstantTeX Picture
\let\picnaturalsize=N
\def\picsize{5.0in}
\def\picfilename{fpi2.eps}
%If you do not have the picture file add:
%\let\nopictures=Y
%to the beginning of the file.
\ifx\nopictures Y\else{\ifx\epsfloaded Y\else \fi
\global\let\epsfloaded=Y
\centerline{\ifx\picnaturalsize N\epsfxsize \picsize\fi\epsfbox{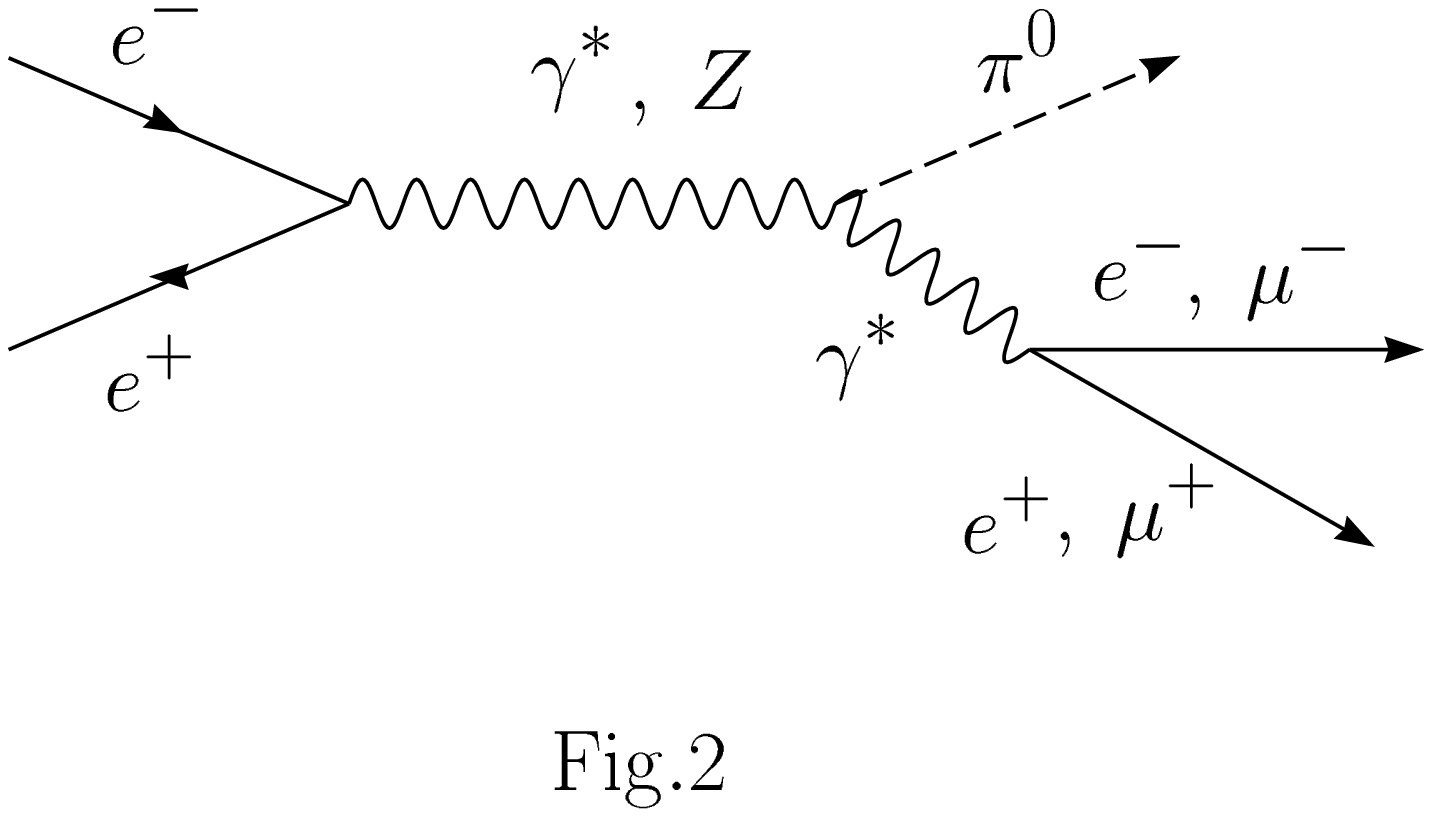}}}\fi
%%End InstantTeX Picture

\vskip 2in
 
%%Begin InstantTeX Picture
\let\picnaturalsize=N
\def\picsize{3.0in}
\def\picfilename{fpi3.eps}
%If you do not have the picture file add:
%\let\nopictures=Y
%to the beginning of the file.
\ifx\nopictures Y\else{\ifx\epsfloaded Y\else \fi
\global\let\epsfloaded=Y
\centerline{\ifx\picnaturalsize N\epsfxsize \picsize\fi
\epsfbox{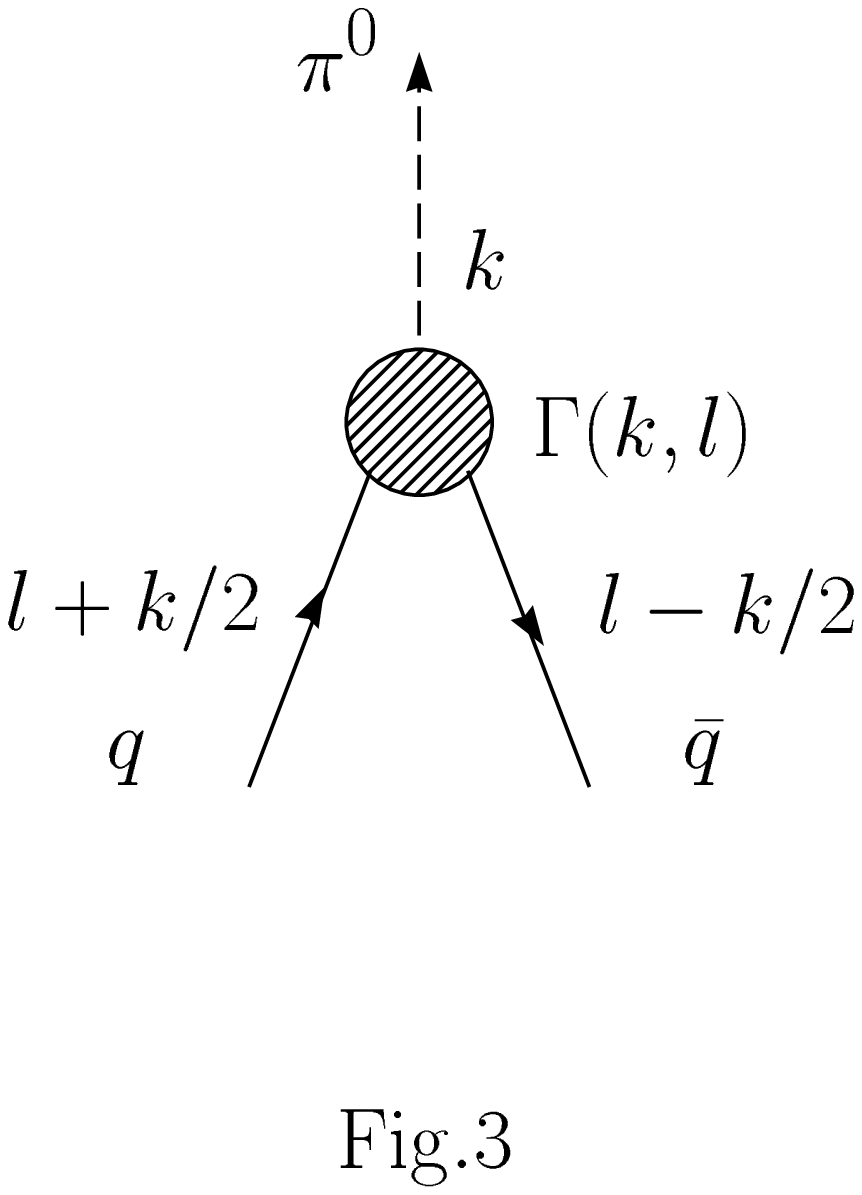}}}\fi
%%End InstantTeX Picture

%%Begin InstantTeX Picture
\let\picnaturalsize=N
\def\picsize{3.0in}
\def\picfilename{fpi4.eps}
%If you do not have the picture file add:
%\let\nopictures=Y
%to the beginning of the file.
\ifx\nopictures Y\else{\ifx\epsfloaded Y\else \fi
\global\let\epsfloaded=Y
\centerline{\ifx\picnaturalsize N\epsfxsize \picsize\fi
\epsfbox{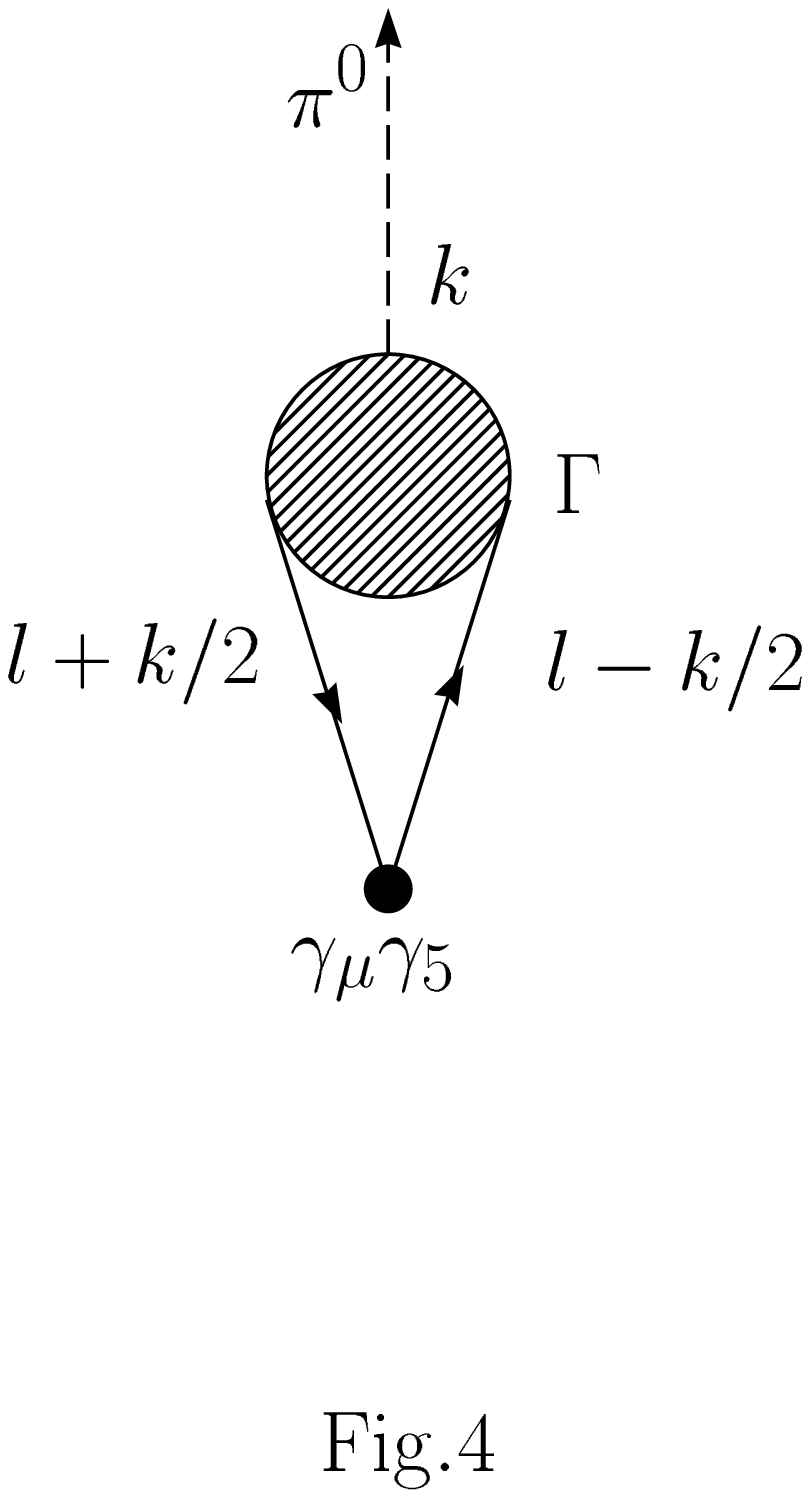}}}\fi
%%End InstantTeX Picture

%%Begin InstantTeX Picture
\let\picnaturalsize=N
\def\picsize{5.0in}
\def\picfilename{fpi5.eps}
%If you do not have the picture file add:
%\let\nopictures=Y
%to the beginning of the file.
\ifx\nopictures Y\else{\ifx\epsfloaded Y\else \fi
\global\let\epsfloaded=Y
\centerline{\ifx\picnaturalsize N\epsfxsize \picsize\fi
\epsfbox{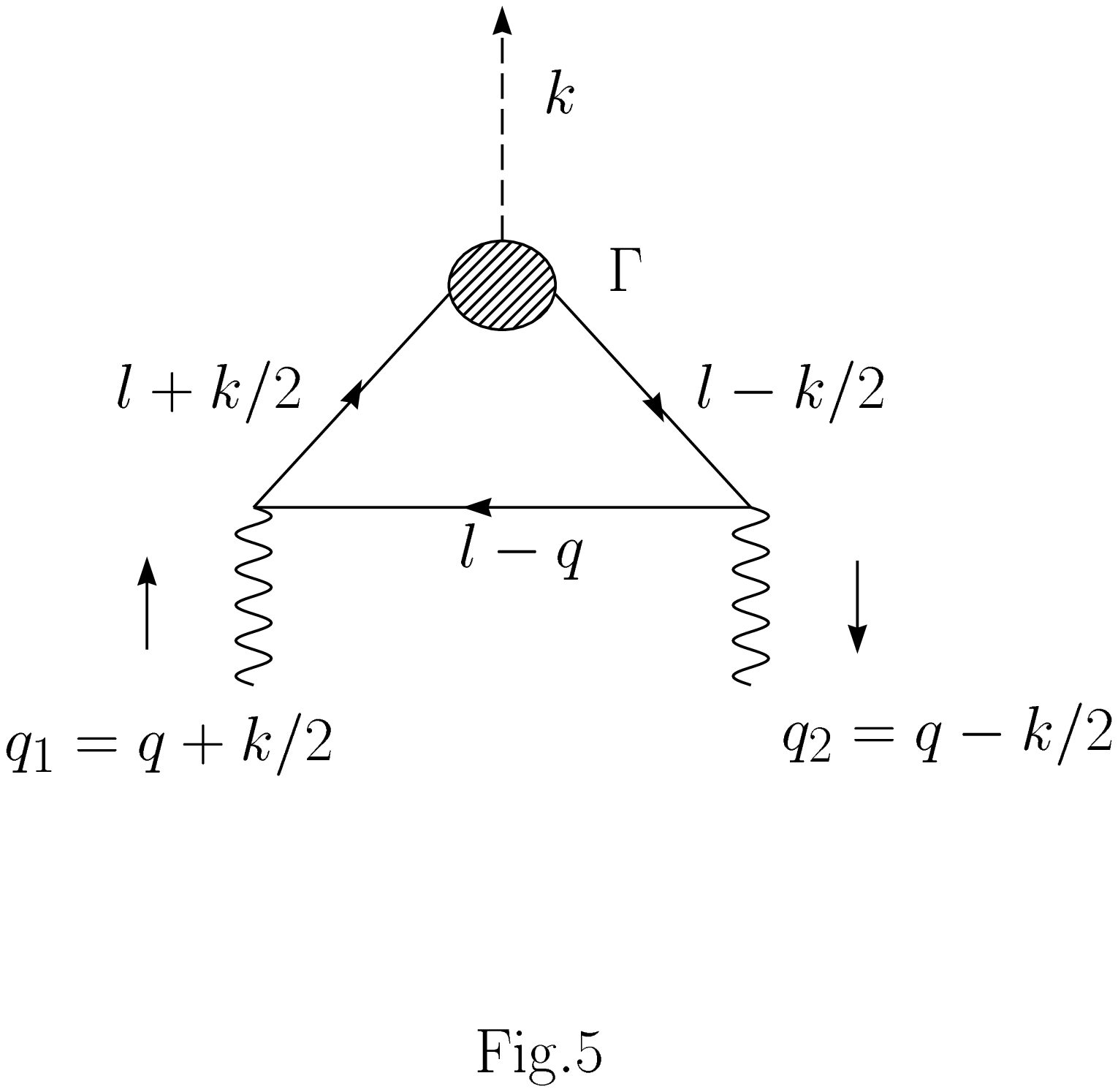}}}\fi
%%End InstantTeX Picture

%%Begin InstantTeX Picture
\let\picnaturalsize=N
\def\picsize{3.0in}
\def\picfilename{fpi6.eps}
%If you do not have the picture file add:
%\let\nopictures=Y
%to the beginning of the file.
\ifx\nopictures Y\else{\ifx\epsfloaded Y\else \fi
\global\let\epsfloaded=Y
\centerline{\ifx\picnaturalsize N\epsfxsize \picsize\fi
\epsfbox{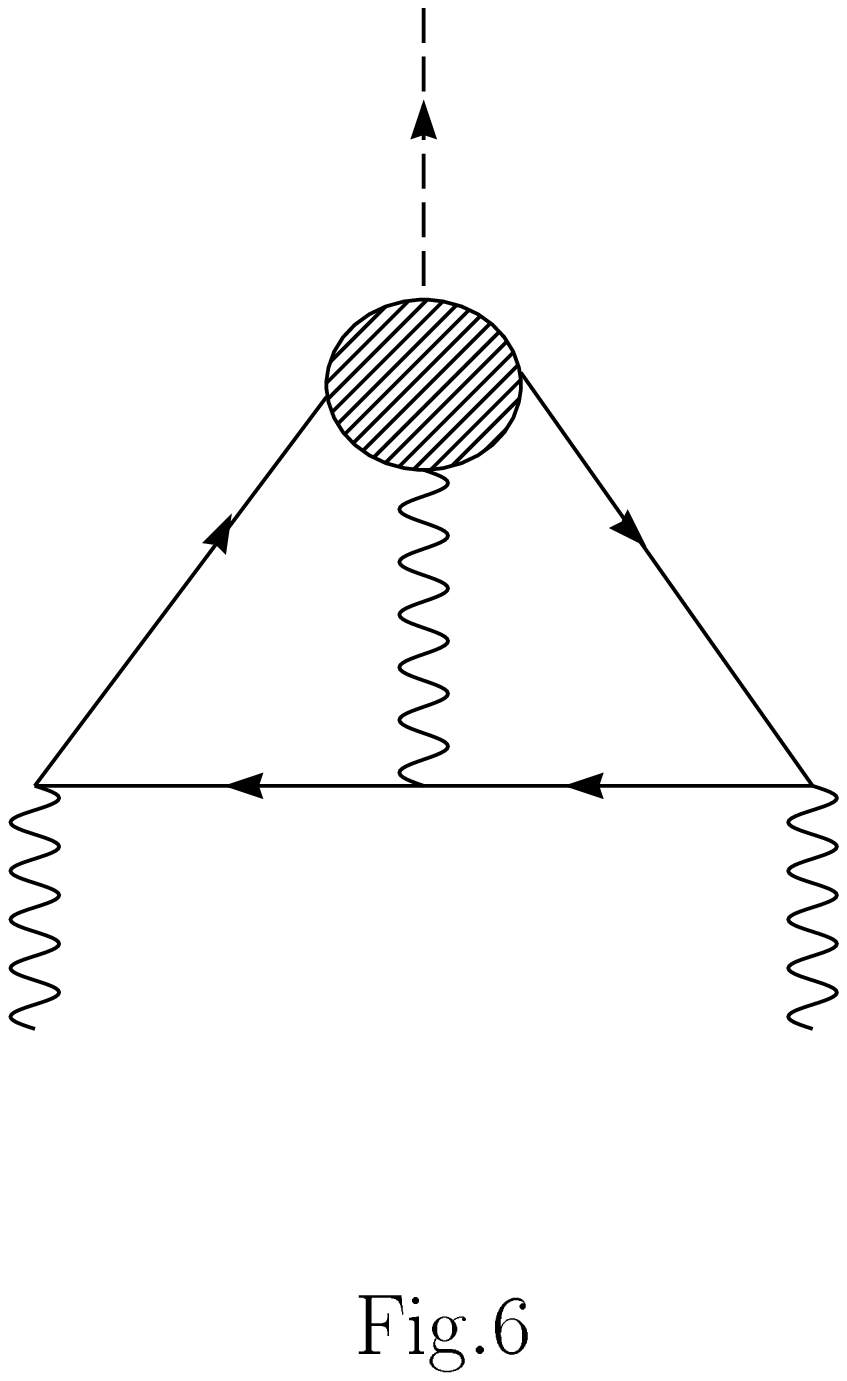}}}\fi
%%End InstantTeX Picture

%%Begin InstantTeX Picture
\let\picnaturalsize=N
\def\picsize{3.0in}
\def\picfilename{fpi7.eps}
%If you do not have the picture file add:
%\let\nopictures=Y
%to the beginning of the file.
\ifx\nopictures Y\else{\ifx\epsfloaded Y\else \fi
\global\let\epsfloaded=Y
\centerline{\ifx\picnaturalsize N\epsfxsize \picsize\fi
\epsfbox{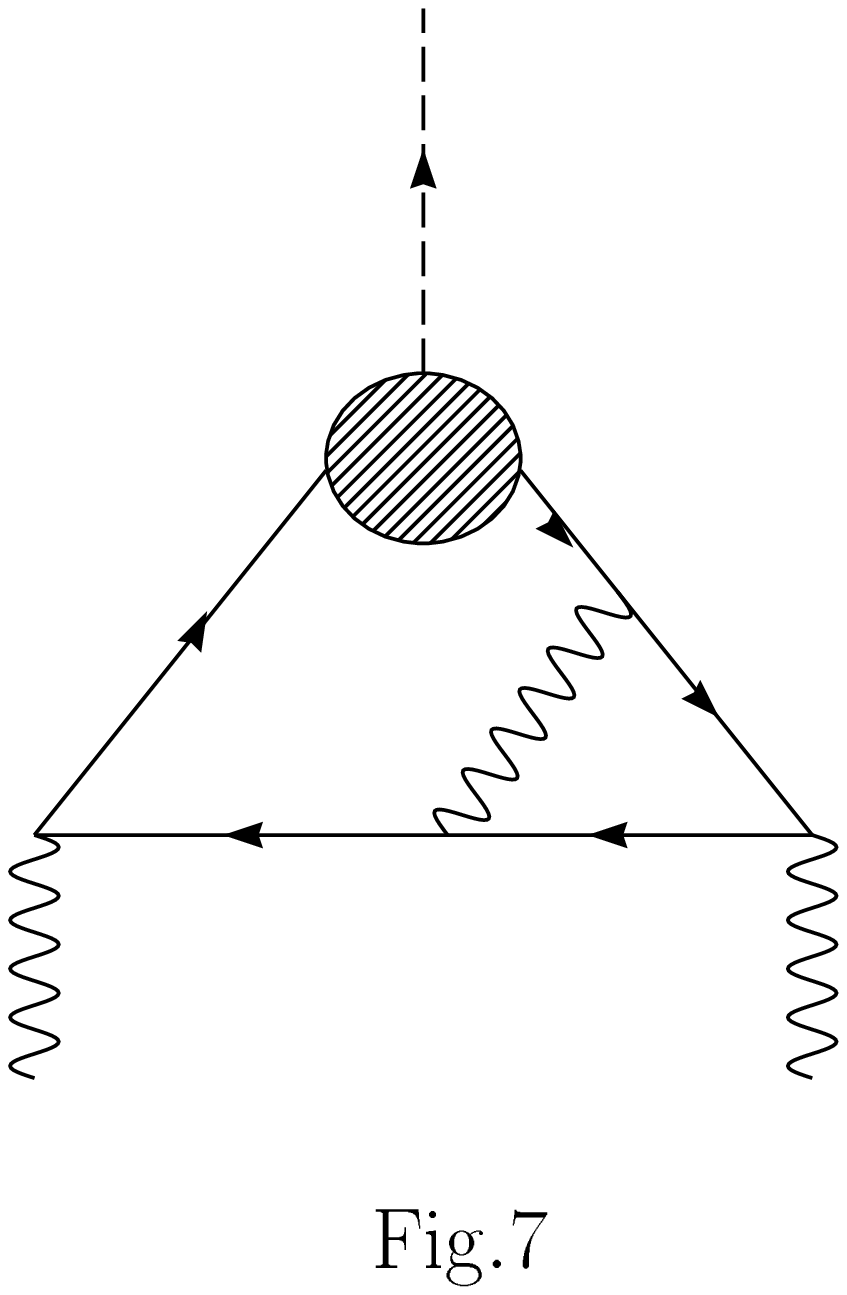}}}\fi
%%End InstantTeX Picture

%%Begin InstantTeX Picture
\let\picnaturalsize=N
\def\picsize{5.0in}
\def\picfilename{fpi8.eps}
%If you do not have the picture file add:
%\let\nopictures=Y
%to the beginning of the file.
\ifx\nopictures Y\else{\ifx\epsfloaded Y\else \fi
\global\let\epsfloaded=Y
\centerline{\ifx\picnaturalsize N\epsfxsize \picsize\fi
\epsfbox{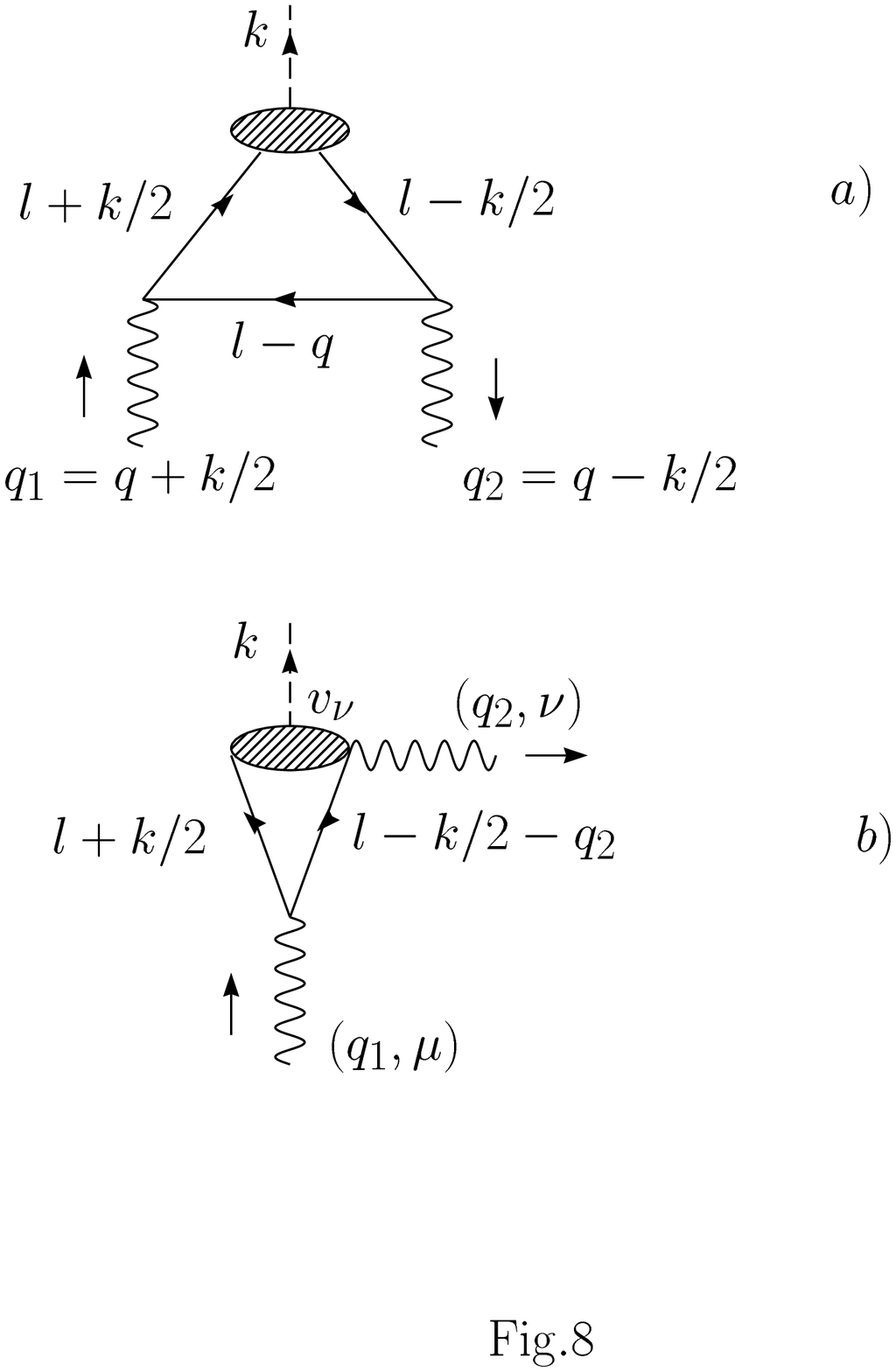}}}\fi
%%End InstantTeX Picture

%%Begin InstantTeX Picture
\let\picnaturalsize=N
\def\picsize{5.0in}
\def\picfilename{fpi9.eps}
%If you do not have the picture file add:
%\let\nopictures=Y
%to the beginning of the file.
\ifx\nopictures Y\else{\ifx\epsfloaded Y\else \fi
\global\let\epsfloaded=Y
\centerline{\ifx\picnaturalsize N\epsfxsize \picsize\fi
\epsfbox{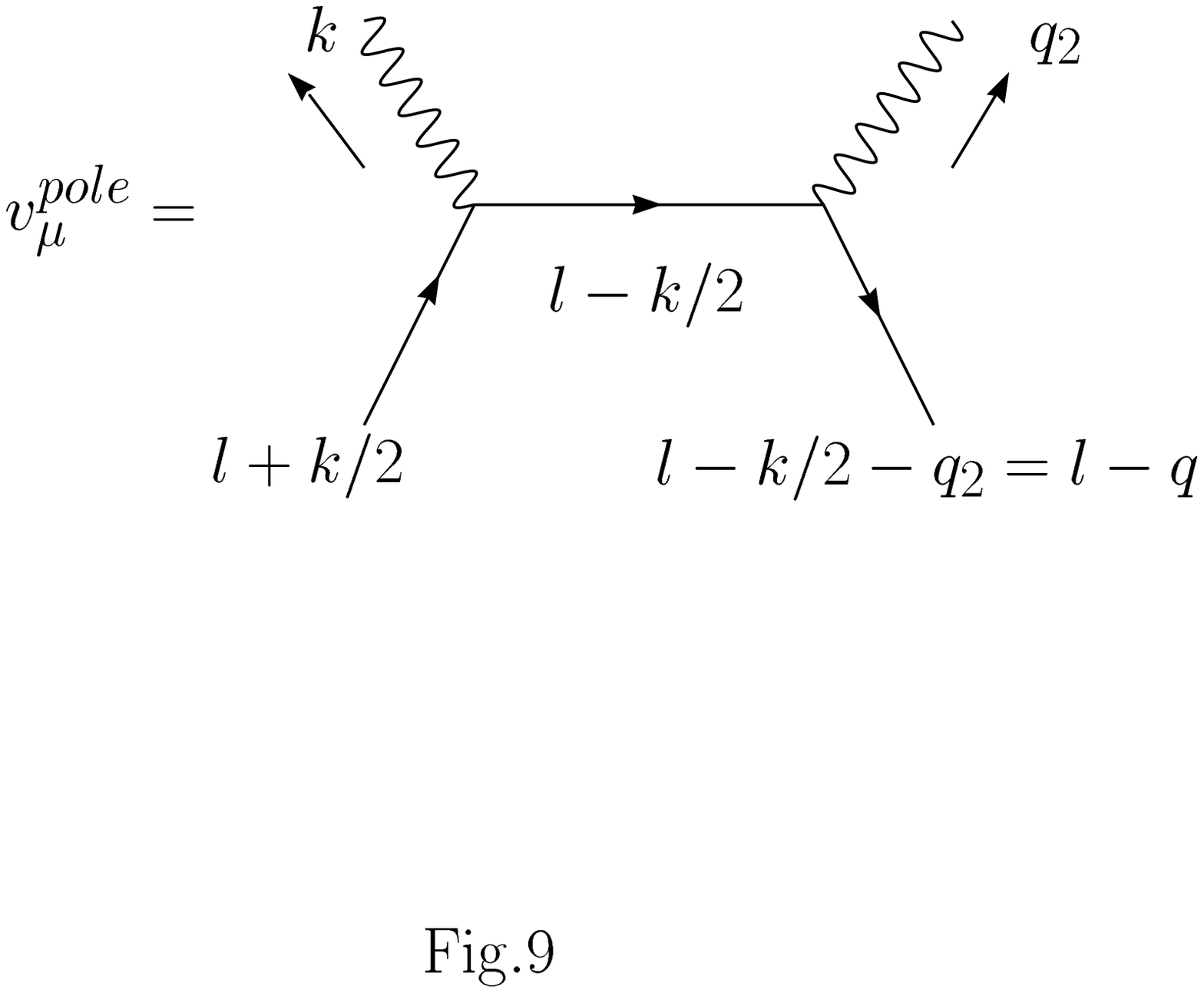}}}\fi
%%End InstantTeX Picture

%%Begin InstantTeX Picture
\let\picnaturalsize=N
\def\picsize{4.0in}
\def\picfilename{fpi10.eps}
%If you do not have the picture file add:
%\let\nopictures=Y
%to the beginning of the file.
\ifx\nopictures Y\else{\ifx\epsfloaded Y\else \fi
\global\let\epsfloaded=Y
\centerline{\ifx\picnaturalsize N\epsfxsize \picsize\fi
\epsfbox{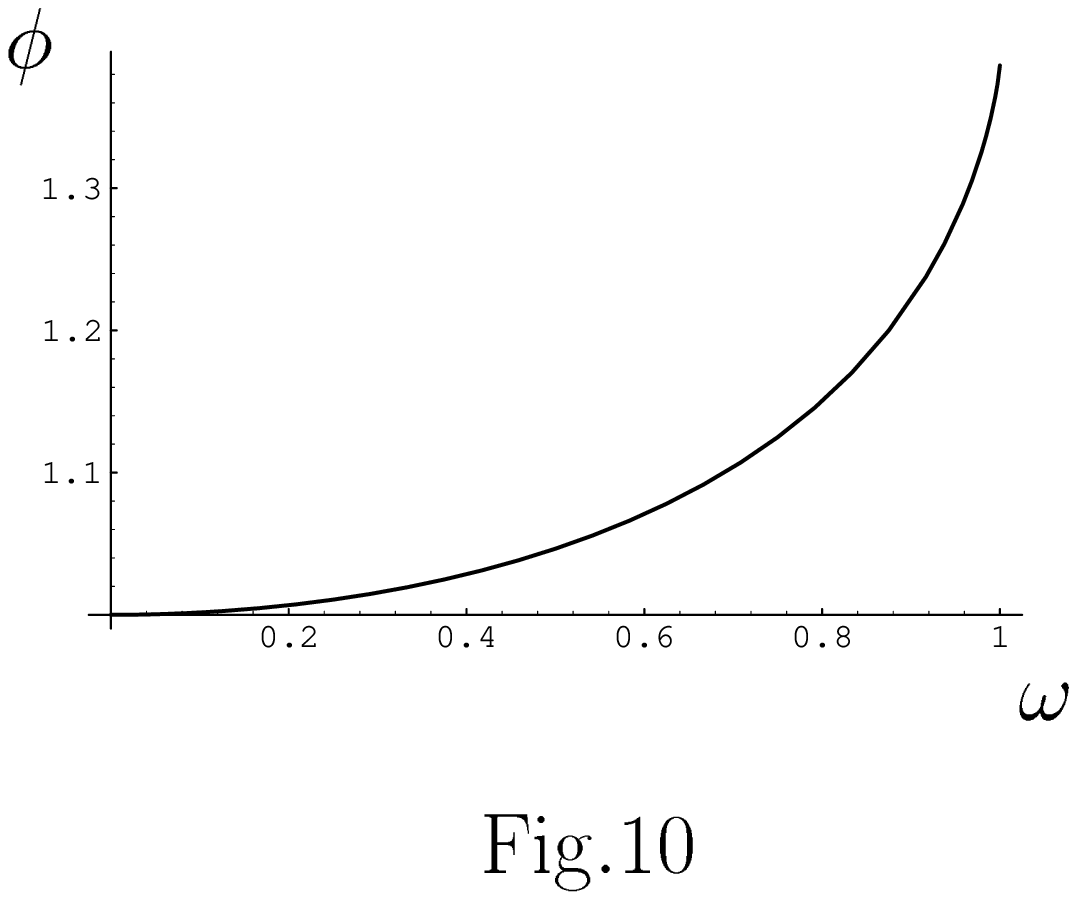}}}\fi
%%End InstantTeX Picture
\vskip 2in

%%Begin InstantTeX Picture
\let\picnaturalsize=N
\def\picsize{4.0in}
\def\picfilename{fpi11.eps}
%If you do not have the picture file add:
%\let\nopictures=Y
%to the beginning of the file.
\ifx\nopictures Y\else{\ifx\epsfloaded Y\else \fi
\global\let\epsfloaded=Y
\centerline{\ifx\picnaturalsize N\epsfxsize \picsize\fi
\epsfbox{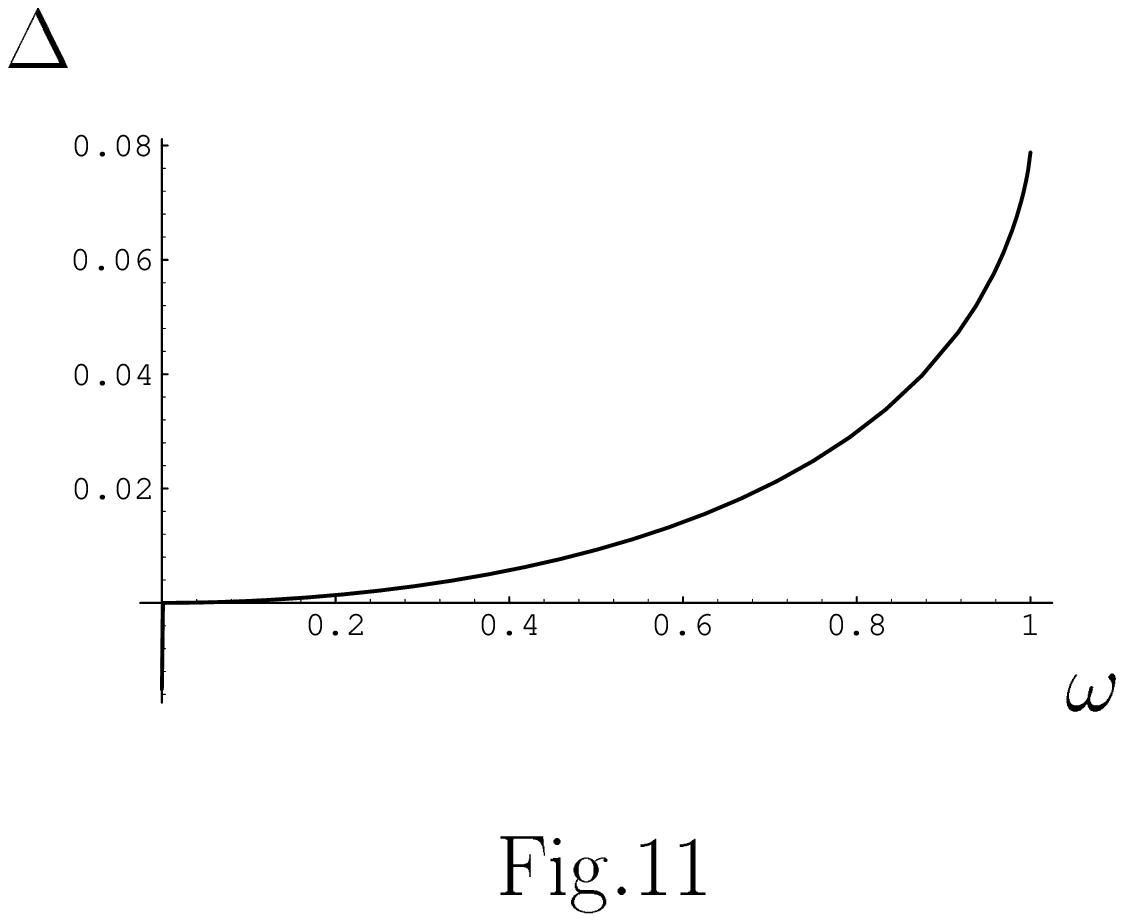}}}\fi
%%End InstantTeX Picture

\end